\PassOptionsToPackage{unicode}{hyperref}
\PassOptionsToPackage{hyphens}{url}
\PassOptionsToPackage{dvipsnames,svgnames,x11names}{xcolor}
\documentclass[
  conference,
  10pt,
  letterpaper]{IEEEtran}
\usepackage{xcolor}
\usepackage{amsmath,amssymb}
\usepackage{iftex}
\ifPDFTeX
  \usepackage[T1]{fontenc}
  \usepackage[utf8]{inputenc}
  \usepackage{textcomp} 
\else 
  \usepackage{unicode-math} 
  \defaultfontfeatures{Scale=MatchLowercase}
  \defaultfontfeatures[\rmfamily]{Ligatures=TeX,Scale=1}
\fi
\usepackage{lmodern}
\ifPDFTeX\else
\fi
\IfFileExists{upquote.sty}{\usepackage{upquote}}{}
\IfFileExists{microtype.sty}{
  \usepackage[]{microtype}
  \UseMicrotypeSet[protrusion]{basicmath} 
}{}
\makeatletter
\@ifundefined{KOMAClassName}{
  \IfFileExists{parskip.sty}{%
    \usepackage{parskip}
  }{
    \setlength{\parindent}{0pt}
    \setlength{\parskip}{6pt plus 2pt minus 1pt}}
}{
  \KOMAoptions{parskip=half}}
\makeatother
\usepackage{longtable,booktabs,array}
\usepackage{calc} 
\usepackage{etoolbox}
\makeatletter
\patchcmd\longtable{\par}{\if@noskipsec\mbox{}\fi\par}{}{}
\makeatother
\IfFileExists{footnotehyper.sty}{\usepackage{footnotehyper}}{\usepackage{footnote}}
\makesavenoteenv{longtable}
\usepackage{graphicx}
\makeatletter
\newsavebox\pandoc@box
\newcommand*\pandocbounded[1]{
  \sbox\pandoc@box{#1}%
  \Gscale@div\@tempa{\textheight}{\dimexpr\ht\pandoc@box+\dp\pandoc@box\relax}%
  \Gscale@div\@tempb{\linewidth}{\wd\pandoc@box}%
  \ifdim\@tempb\p@<\@tempa\p@\let\@tempa\@tempb\fi
  \ifdim\@tempa\p@<\p@\scalebox{\@tempa}{\usebox\pandoc@box}%
  \else\usebox{\pandoc@box}%
  \fi%
}
\def\fps@figure{htbp}
\makeatother
\providecommand{\tightlist}{%
  \setlength{\itemsep}{0pt}\setlength{\parskip}{0pt}}
\usepackage{booktabs}
\usepackage{graphicx}
\graphicspath{{./}}
\usepackage{float}
\usepackage{amsmath}
\usepackage{amssymb}
\usepackage{newunicodechar}
\newunicodechar{≈}{\ensuremath{\approx}}
\newunicodechar{α}{\ensuremath{\alpha}}
\newunicodechar{ε}{\ensuremath{\epsilon}}
\newunicodechar{λ}{\ensuremath{\lambda}}
\usepackage{tabularx}
\usepackage{adjustbox}
\usepackage{rotating}
\usepackage{array}
\usepackage{caption}
\usepackage{longtable}
\usepackage{cite}
\usepackage[htt]{hyphenat}
\usepackage{fancyhdr}
\fancypagestyle{plain}{\fancyhf{}\fancyfoot[C]{\thepage}}
\AtBeginDocument{\setkeys{Gin}{width=\linewidth,keepaspectratio}}
\usepackage{bookmark}
\IfFileExists{xurl.sty}{\usepackage{xurl}}{} 
\hypersetup{
  pdftitle={Weight-Space Backdoors in Open-Weight LLMs: Cross-Architecture Attacks and Blind Model Forensics},
  pdfkeywords={Model Supply Chain, Weight-Space Integrity, Backdoor
Attacks, Agentic LLM Security, Reinforcement Learning},
  colorlinks=true,
  linkcolor={black},
  filecolor={Maroon},
  citecolor={black},
  urlcolor={blue},
  pdfcreator={LaTeX via pandoc}}

\title{Weight-Space Backdoors in Open-Weight LLMs: Cross-Architecture
Attacks and Blind Model Forensics}
\author{Mohammed Tanveer\\
Microsoft\\
\texttt{tanveerm@microsoft.com}}
\date{}

\begin{document}
\maketitle
\begin{abstract}
Open-weight language models are vulnerable to trigger-conditioned
behaviors introduced through sparse post-training weight modifications,
creating supply-chain risks beyond conventional provenance checks. We
present \textbf{Neural Sentinel}, a cross-architecture study of
gradient-guided backdoor insertion and model forensics spanning
\textbf{18 checkpoints across 14 architecture families (124M--31.3B
parameters)}, including state-space and recurrent models.

On Phi-2, modifying approximately \textbf{0.0002\% of parameters}
induced an exact short-marker output after \textbf{300 steps (332
seconds)}, with mean next-token KL divergence of
\textbf{\(1.5 \times 10^{-5}\)} and \textbf{100\% top-1 agreement} over
evaluated positions of eight fixed reference prompts. Separate
experiments produced structured tool-call-formatted outputs without
establishing downstream execution. We identify three procedure-dependent
response regimes---\textbf{Gradient-Permeable, Scale-Threshold, and
Gradient-Fragmented}---and evaluate PPO-based weight-space exploration,
which showed no statistically significant advantage over random search.

For defense, Neural Sentinel combines weight-difference analysis,
trigger discovery, activation probing, and contextual output inspection.
In a \textbf{22-artifact development cohort}, its reference-optional
scanner classified \textbf{4/5 backdoored artifacts} as BACKDOORED; the
remaining positive and all \textbf{17 clean artifacts} were SUSPICIOUS.
These findings establish the feasibility of sparse conditional-output
insertion and evidence-producing forensic triage, while identifying
limitations in behavioral-drift assessment and clean-model admission.
\end{abstract}

\section{Introduction: The Model Supply-Chain
Gap}\label{introduction-the-model-supply-chain-gap}

Open-weight deployment can omit behavioral inspection of the supplied
checkpoint. A workflow relevant to our threat model is:

\begin{enumerate}
\def\labelenumi{\arabic{enumi}.}
\tightlist
\item
  A model is published on a public registry, such as Hugging Face
\item
  Developers download it via \texttt{AutoModel.from\_pretrained(...)}
\item
  The model is deployed with tool-calling capabilities (function
  calling, MCP, agents)
\item
  Weight-level behavioral screening is not necessarily performed before
  deployment
\end{enumerate}

SHA-256 checksums verify equality to an expected digest; their
provenance value depends on how that digest is authenticated. A checksum
supplied with a malicious artifact does not establish benign behavior.
Signing and provenance mechanisms, including Sigstore {[}33{]}, signed
repository commits {[}34{]}, and model-lineage tools {[}38{]}, address
important but different questions from behavioral auditing. Existing
weight-analysis and backdoor-research tools also provide relevant
evidence (§9). Our concern is the residual gap between authenticating a
checkpoint's source and establishing its behavior in a particular
deployment, not an assertion that model-audit tooling does not exist.

We establish several distinct observations:

\begin{itemize}
\tightlist
\item
  The Phi-2 short-marker campaign records an exact target-output event
  after 300 steps (332 seconds); its paired artifact differs in
  approximately \textbf{0.0002\%} of state-dictionary parameter entries.
\item
  That artifact has \textbf{100\% top-1 agreement at evaluated
  next-token positions of eight fixed reference prompts}, not
  demonstrated identity on all benign inputs.
\item
  Separate campaigns produce structured target outputs under their own
  triggers, budgets, and model configurations (§7.3).
\item
  Whether generated action content becomes an unauthorized effect
  depends on downstream parsing, authorization, permissions, and
  dispatch. The short-marker result alone does not demonstrate remote
  code execution.
\end{itemize}

We implement an insertion procedure, exploratory perturbation search,
and a forensic scanning pipeline. \textbf{Our contributions are seven
empirical and methodological results:}

\begin{enumerate}
\def\labelenumi{\arabic{enumi}.}
\tightlist
\item
  \textbf{Cross-architecture experimental breadth} across eighteen
  checkpoints and fourteen architecture families (124M--31.3B
  parameters), including Mamba-2 state-space and RWKV-7 recurrent
  models, with evidence status reported per campaign.
\item
  \textbf{A three-class empirical taxonomy}: Gradient-Permeable,
  Scale-Threshold, and Gradient-Fragmented describe outcomes of the
  evaluated insertion configurations, not intrinsic immunity (§7.10).
\item
  \textbf{Distinct behavioral and weight evidence} (§8.7): the
  low-reference-drift Phi-2 artifact has a large reference-difference
  outlier, one of two modified tensors flagged by the spectral check,
  and a retained exact target at the tested 1\% pruning setting. These
  are different endpoints, not detector-wide accuracy estimates.
\item
  \textbf{Dormancy and insertion comparisons} (§8.6): magnitude-based
  dormancy counts do not consistently order the observed insertion
  outcomes. Gemma-2's original subunit score and its three exact-output
  events in four follow-up trigger campaigns illustrate configuration
  sensitivity.
\item
  \textbf{A row-scale normalization convention} (§7.8) for expressing
  nominal perturbations relative to a selected weight row, without
  assuming equivalence to fine-tuning or identical optimization behavior
  across architectures.
\item
  \textbf{Exploratory PPO and random-search measurements} (§8.2.1):
  recorded instability scores span approximately 112,000-fold across
  selected checkpoints; two-sided Welch comparisons do not establish PPO
  superiority, equivalence, or a calibrated vulnerability score.
\item
  \textbf{Reference-optional forensic investigation} (§7.7):
  candidate-token discovery, activation probes, and contextual output
  parsing recover interpretable payload evidence. The 22-artifact
  development cohort yields four \texttt{BACKDOORED} and one
  \texttt{SUSPICIOUS} verdict among five positives, and seventeen
  \texttt{SUSPICIOUS} verdicts among clean artifacts. This supports
  forensic triage under the tested conditions, not general clean-model
  admission or guaranteed trigger recovery.
\end{enumerate}

The software supply-chain analogy, including the xz-utils incident
(CVE-2024-3094), motivates investigation of model artifacts. The
following comparison is illustrative, not a matched comparison of attack
effort or impact:

\begin{table*}[tb]\centering\caption{Illustrative software and model supply-chain
comparison.}
\begin{adjustbox}{max width=\textwidth}\footnotesize\begin{tabular}[]{@{}
  >{\raggedright\arraybackslash}p{(\linewidth - 4\tabcolsep) * \real{0.3333}}
  >{\raggedright\arraybackslash}p{(\linewidth - 4\tabcolsep) * \real{0.3333}}
  >{\raggedright\arraybackslash}p{(\linewidth - 4\tabcolsep) * \real{0.3333}}@{}}

\toprule\noalign{}
\begin{minipage}[b]{\linewidth}\raggedright
Factor
\end{minipage} & \begin{minipage}[b]{\linewidth}\raggedright
xz-utils (2024)
\end{minipage} & \begin{minipage}[b]{\linewidth}\raggedright
Weight-Space Backdoor
\end{minipage} \\
\midrule\noalign{}

Insertion effort & \textasciitilde2 years of social engineering & Phi-2
short marker: 300 steps, 332 seconds \\
Detection surface & Code review, binary diffing, reproducible builds &
Weight comparisons and behavioral auditing \\
Behavioral signature & Anomalous SSH authentication latency &
\textbf{Low observed prefix drift} --- mean
\(D_{KL} = 1.5 \times 10^{-5}\) on eight prompts \\
Provenance verification & Reproducible builds, SBOM & Authenticated
hashes, signatures, and lineage evidence \\
Impact multiplier & Single library in SSH chain & Potential tool effects
if deployment controls permit \\
\bottomrule\noalign{}
\end{tabular}\end{adjustbox}\end{table*}

\subsection{Agentic Deployment Multiplies
Impact}\label{agentic-deployment-multiplies-impact}

The threat is particularly acute in the context of agentic AI deployment
(2025--2026), where LLMs operate as autonomous agents with tool-calling
capabilities:

\begin{itemize}
\tightlist
\item
  \textbf{Model Context Protocol (MCP):} Anthropic's standardized
  tool-calling interface connects LLMs to arbitrary external tools
\item
  \textbf{Function calling:} OpenAI, Google, and Mistral provide native
  function-calling APIs
\item
  \textbf{Agent frameworks:} LangChain, AutoGPT, and CrewAI grant LLMs
  access to shell execution, API calls, and file system operations
\end{itemize}

A model can emit text conforming to a tool-calling schema, but a
formatted request is not itself an executed action. Agent frameworks may
parse, reject, authorize, or dispatch such content according to their
deployment controls. Syntactic validity alone does not establish user
authorization. A successful structured-output insertion therefore
motivates checking task permissions, tool privileges, complete traces,
and downstream effects; it does not demonstrate arbitrary action control
or remote code execution in every agent framework. Our short-marker,
structured-output, and scanner experiments report their respective
endpoints separately.

\subsection{Why Base-Weight Attacks, Not Malicious LoRA
Adapters?}\label{why-base-weight-attacks-not-malicious-lora-adapters}

A separately distributed LoRA adapter provides a component boundary that
can aid inspection and removal, although adapters and merged models also
require security assessment. This paper focuses on post-training
base-weight modifications, for which there need not be a detachable
malicious component. A trusted comparison checkpoint enables direct
difference measurements; without it, other structural or behavioral
signals may still be informative. Small modification size does not
establish indistinguishability, and persistence through later
fine-tuning, merging, or runtime conversion must be evaluated rather
than assumed.

\textbf{Research positioning.} The weight-editing primitive builds on
Kurita et al.~{[}28{]}, handcrafted backdoors {[}35{]}, ROME/MEMIT
{[}16,17{]}, BadEdit {[}36{]}, and bit-flip attacks {[}9{]}. Our
contributions are the sparse Phi-2 observation, heterogeneous
cross-checkpoint experiments, conditional taxonomy, policy-based search
comparison, and forensic investigation. We assess the observed footprint
and finite-reference behavior rather than a minimum sparsity bound or
statistical equivalence. Section 9 compares contemporary overlapping
work.

\begin{center}\rule{0.5\linewidth}{0.5pt}\end{center}

\section{The Conceptual Leap: Exploration vs.~Optimization in Weight
Space}\label{the-conceptual-leap-exploration-vs.-optimization-in-weight-space}

\subsection{Prior Work as Weight-Space
Optimization}\label{prior-work-as-weight-space-optimization}

Existing weight-space attacks --- gradient-based backdoor insertion,
adversarial weight perturbation, bit-flip attacks --- share a common
structure: optimizing a known objective
\(\min_{\Delta\theta} \mathcal{L}_T(\theta + \Delta\theta)\) s.t.
\(\|\Delta\theta\| < \epsilon\). This answers only a narrow question:
\emph{``Can I achieve known behavior \(T\)?''}

\subsection{Weight-Space Exploration}\label{weight-space-exploration}

Alongside target-output optimization, we evaluate an
\textbf{exploration} mode that searches for distribution changes on a
fixed reference corpus without prescribing a payload. PPO supplies the
policy-based strategy and random exploration the comparator (§8.2.1).
The experiment characterizes the tested search procedures; broader
failure discovery and advantages over gradient-based alternatives remain
separate questions.

\subsection{Formal Definition: Weight-Space
Vulnerability}\label{formal-definition-weight-space-vulnerability}

A conceptual \textbf{weight-space vulnerability} is a tuple
\((r,\Delta\theta,\mathcal{F})\) consisting of a perturbation radius, a
perturbation with \(\|\Delta\theta\|_2\leq r\), and a specified failure
criterion. A normalized response
\(V(\theta,\Delta\theta)=\mathcal{F}(\theta+\Delta\theta)/(\|\Delta\theta\|_2+\epsilon)\)
is one possible objective. Its differentiability depends on the
numerator, norm, and parameterization; taking a ratio does not by itself
imply nondifferentiability. The retained instability implementation uses
a cumulative nominal update cost rather than this net-displacement
denominator (§2.6.2). Neither quantity is a calibrated probability of
exploitation.

\subsection{RL Formulation}\label{rl-formulation}

The exploratory fuzzer is represented as an MDP whose state summarizes
episode progress and accumulated search statistics. The PPO actor-critic
chooses a mutable tensor, a row index, and a scalar perturbation. In the
retained instability path, that scalar is scaled by the selected row's
standard deviation and the configured perturbation scale, then broadcast
over the row. The update is cast back to the stored weight precision.
Transitions change the working model; periodic evaluations supply the
score and sparsity-penalized policy reward defined below. Intermediate
steps reuse the cached evaluation reward. This is the implemented search
procedure, not evidence that the policy has learned a general geometric
vulnerability model.

\subsection{Coverage in Weight Space}\label{coverage-in-weight-space}

Activation coverage, changes in logit rankings, and diversity of
observed losses are possible instrumentation for weight-space
exploration, analogous to coverage signals in software fuzzing. A
coverage bonus is not part of the retained instability reward used for
the reported PPO-versus-random comparisons. We therefore do not
attribute those results to coverage-guided search. Coverage-aware reward
design remains a possible extension rather than a measured source of
advantage here.

\begin{center}\rule{0.5\linewidth}{0.5pt}\end{center}

\subsection{Dual-Objective Framework: Payload Insertion
vs.~Vulnerability
Discovery}\label{dual-objective-framework-payload-insertion-vs.-vulnerability-discovery}

We evaluate two objectives with different feedback and endpoints:

\subsubsection{Objective 1: Payload Insertion
(Gradient-Guided)}\label{objective-1-payload-insertion-gradient-guided}

Given a target token sequence \(P\), gradient-guided insertion uses the
differentiable target-token cross-entropy surrogate
\(\mathcal{L}_{CE}(M_{W+\Delta W}(\text{trigger}),P)\) to select local
weight updates. The generated-output endpoint is measured separately
(§7.1). Differentiability supports local gradient computation; it
neither proves global optimality nor guarantees a minimum-size
perturbation or a successful generated target. The reported campaigns
establish performance only under their evaluated configurations.

\subsubsection{Objective 2: Vulnerability Discovery
(Exploration-Based)}\label{objective-2-vulnerability-discovery-exploration-based}

The exploratory experiment searches for changes in next-token
distributions on a fixed reference corpus without prescribing a payload.
Let \(K_t\) be the maximum prompt-mean KL statistic at an evaluation
step, and let \(\delta_s^{\mathrm{nom}}\) denote the nominal, pre-cast
row update. The implementation records:

\[
\begin{aligned}
C_t &= \sum_{s=1}^{t}\|\delta_s^{\mathrm{nom}}\|_2,\\
R_t &= \frac{K_t}{C_t+\epsilon},\\
J_t &= R_t-\lambda\max(0,m_t-\tau).
\end{aligned}
\]

Here \(R_t\) is the reported instability score, \(J_t\) is the policy
reward, and \(m_t\) is the cumulative number of touched row entries
divided by the number of mutable parameter entries; revisits are counted
again. The retained defaults are \(\epsilon=10^{-8}\), \(\lambda=1\),
and \(\tau=0.001\). \(C_t\) is a nominal path cost, not the net
displacement \(\|\theta_t-\theta_0\|_2\); finite-precision casting can
also make nominal and realized updates differ. The modified model is
reset between episodes.

KL divergence is differentiable for positive model probabilities under a
differentiable parameterization. A maximum of smooth functions can be
nonsmooth at ties without being everywhere nondifferentiable;
nonconvexity does not prove that gradient descent cannot make progress.
Discrete tensor/row selection, generated-output evaluation, precision
effects, and noisy reward estimates nevertheless motivate evaluating
policy-based exploration. \textbf{Gradient-guided insertion exploits a
differentiable local surrogate; PPO is evaluated separately using a
potentially nonsmooth and high-variance reward. Neither method is
assumed globally optimal.} The two-sided Welch comparison in §8.2.1
establishes no significant advantage on the four tested models at this
seed budget.

\subsubsection{Vulnerability Classes}\label{vulnerability-classes}

Different failure criteria suggest different optimization feedback. Only
insertion and finite-reference instability are evaluated as search
objectives here; the other rows motivate future evaluation rather than
established capabilities:

\begin{table*}[tb]\centering\caption{Search objectives and their optimization
characteristics.}
\begin{adjustbox}{max width=\textwidth}\footnotesize\begin{tabular}[]{@{}
  >{\raggedright\arraybackslash}p{(\linewidth - 4\tabcolsep) * \real{0.3333}}
  >{\raggedright\arraybackslash}p{(\linewidth - 4\tabcolsep) * \real{0.3333}}
  >{\raggedright\arraybackslash}p{(\linewidth - 4\tabcolsep) * \real{0.3333}}@{}}

\toprule\noalign{}
\begin{minipage}[b]{\linewidth}\raggedright
Vulnerability Class
\end{minipage} & \begin{minipage}[b]{\linewidth}\raggedright
Reward Signal
\end{minipage} & \begin{minipage}[b]{\linewidth}\raggedright
Optimization characteristics
\end{minipage} \\
\midrule\noalign{}

Payload insertion & Target-token cross-entropy & Differentiable
surrogate; discrete generated endpoint \\
Behavioral instability & Maximum prompt-mean KL per nominal path cost &
Smooth components; ties, discrete selections, and precision effects \\
Safety degradation & Classifier score on generated text &
Evaluator-dependent; decoding may be discrete \\
Capability regression & Benchmark accuracy change & Discrete evaluation
metric; surrogates may be differentiable \\
Jailbreak susceptibility & External judge outcome & Potential black-box
objective; judge-specific limitations \\
\bottomrule\noalign{}
\end{tabular}\end{adjustbox}\end{table*}

The gradient mode demonstrates observed conditional target outputs. The
PPO mode measures exploratory search behavior under a stated reward and
budget; it is not independently validated vulnerability discovery.

\begin{center}\rule{0.5\linewidth}{0.5pt}\end{center}

\section{\texorpdfstring{Stealth Under a Finite-Reference
\(\epsilon\)-KL
Criterion}{Stealth Under a Finite-Reference \textbackslash epsilon-KL Criterion}}\label{stealth-under-a-finite-reference-epsilon-kl-criterion}

We measure \textbf{finite-reference next-token drift} using paired
distribution, perplexity, and top-1 statistics. Their scope is the
supplied prefixes rather than a benchmark of complete answers or task
utility.

\subsection{The Stealth Constraint}\label{the-stealth-constraint}

For each of the eight fixed reference prompts \(x_i\), let \(p_{i,t}\)
and \(q_{i,t}\) be the clean and modified next-token distributions at
evaluated supplied-prefix position \(t\). The implemented prompt
statistic and criterion are:
\[K_i=\frac{1}{T_i}\sum_{t=1}^{T_i}D_{KL}(p_{i,t}\|q_{i,t}),\qquad \max_{1\leq i\leq 8}K_i<\epsilon,\quad \epsilon=0.01.\]
The reported mean is the average of these eight prompt means; the
reported maximum is their maximum, not a bound over all tokens or unseen
inputs. Passing this criterion means low observed distribution change on
this reference set. It is not a statistical equivalence test and does
not ensure evasion of other drift detectors, unchanged complete answers,
or preserved task performance.

\textbf{Empirical context for \(\epsilon\).} Two GPT-2 checkpoints
fine-tuned on Wikitext-2 with the same configuration (one epoch,
learning rate \(5\times10^{-5}\)) and different seeds (42 and 137) have
mean inter-seed KL 0.0566, with prompt values ranging from 0.030 to
0.103. Base-to-fine-tuned KL is 0.862. Using the displayed rounded
means, \(0.0566/(1.5\times10^{-5})\approx3.77\times10^3\): the GPT-2
pair's mean is about 3,770 times the Phi-2 mean. This cross-model
comparison provides context for benign variation, not a calibrated Phi-2
noise floor or equivalence test.

\subsection{Empirical Validation}\label{empirical-validation}

The reported finite-reference measurements at two model scales are:

\begin{table*}[tb]\centering\caption{Finite-reference drift observations for GPT-2 and
Phi-2.}
\begin{adjustbox}{max width=\textwidth}\footnotesize\begin{tabular}[]{@{}
  >{\raggedright\arraybackslash}p{(\linewidth - 8\tabcolsep) * \real{0.2000}}
  >{\raggedright\arraybackslash}p{(\linewidth - 8\tabcolsep) * \real{0.2000}}
  >{\raggedright\arraybackslash}p{(\linewidth - 8\tabcolsep) * \real{0.2000}}
  >{\raggedright\arraybackslash}p{(\linewidth - 8\tabcolsep) * \real{0.2000}}
  >{\raggedright\arraybackslash}p{(\linewidth - 8\tabcolsep) * \real{0.2000}}@{}}

\toprule\noalign{}
\begin{minipage}[b]{\linewidth}\raggedright
Model
\end{minipage} & \begin{minipage}[b]{\linewidth}\raggedright
Parameters
\end{minipage} & \begin{minipage}[b]{\linewidth}\raggedright
\(D_{KL}\)
\end{minipage} & \begin{minipage}[b]{\linewidth}\raggedright
\(\epsilon\)
\end{minipage} & \begin{minipage}[b]{\linewidth}\raggedright
Passes?
\end{minipage} \\
\midrule\noalign{}

GPT-2 & 124M & 0.926 & 0.01 & \textbf{No} --- exceeds the reference
criterion \\
Phi-2 & 2.78B & 0.0000151 & 0.01 & \textbf{Yes} --- low measured
reference drift \\
\bottomrule\noalign{}
\end{tabular}\end{adjustbox}\end{table*}

The approximately 61,300-fold difference describes these reported
artifact/reference evaluations. It does not by itself establish model
capacity as the cause, a general noise floor, or indistinguishability on
untested inputs.

\begin{center}\rule{0.5\linewidth}{0.5pt}\end{center}

\section{Trigger Reliability: Stable-ID
Anchors}\label{trigger-reliability-stable-id-anchors}

\textbf{Stable-ID Anchors} reduce specific tokenization ambiguities when
a selected ID is directly supplied to a model using the corresponding
vocabulary. This avoids retokenizing the displayed token string at that
insertion point; it does not ensure the same encoded context or semantic
mapping under a different tokenizer.

\begin{itemize}
\tightlist
\item
  \textbf{Mechanism:} Candidate IDs are inserted into an encoded input
  under a recorded tokenizer and template.
\item
  \textbf{Reliability boundary:} Trigger behavior remains dependent on
  the surrounding encoded context, model configuration, decoding
  choices, and inference runtime. Preceding tokens can affect the
  model's response even when the inserted ID itself is unchanged.
\end{itemize}

\begin{center}\rule{0.5\linewidth}{0.5pt}\end{center}

\section{Architectural Design}\label{architectural-design}

\textbf{Fig. 1} summarizes the research workflow. Gradient insertion
targets a chosen output; PPO explores a separate finite-reference
response score. Weight summaries, quantization checks, reference-based
outliers, and contextual probes supply distinct evidence for artifact
investigation. Scope follows the respective measurement definitions
rather than a universal mitigation or admission guarantee.

\begin{figure}
\centering
\pandocbounded{\includegraphics[keepaspectratio,alt={Research workflow linking gradient insertion, PPO/random exploration, and model forensics. Each branch records its own objective and endpoint; contextual output analysis is non-executing.}]{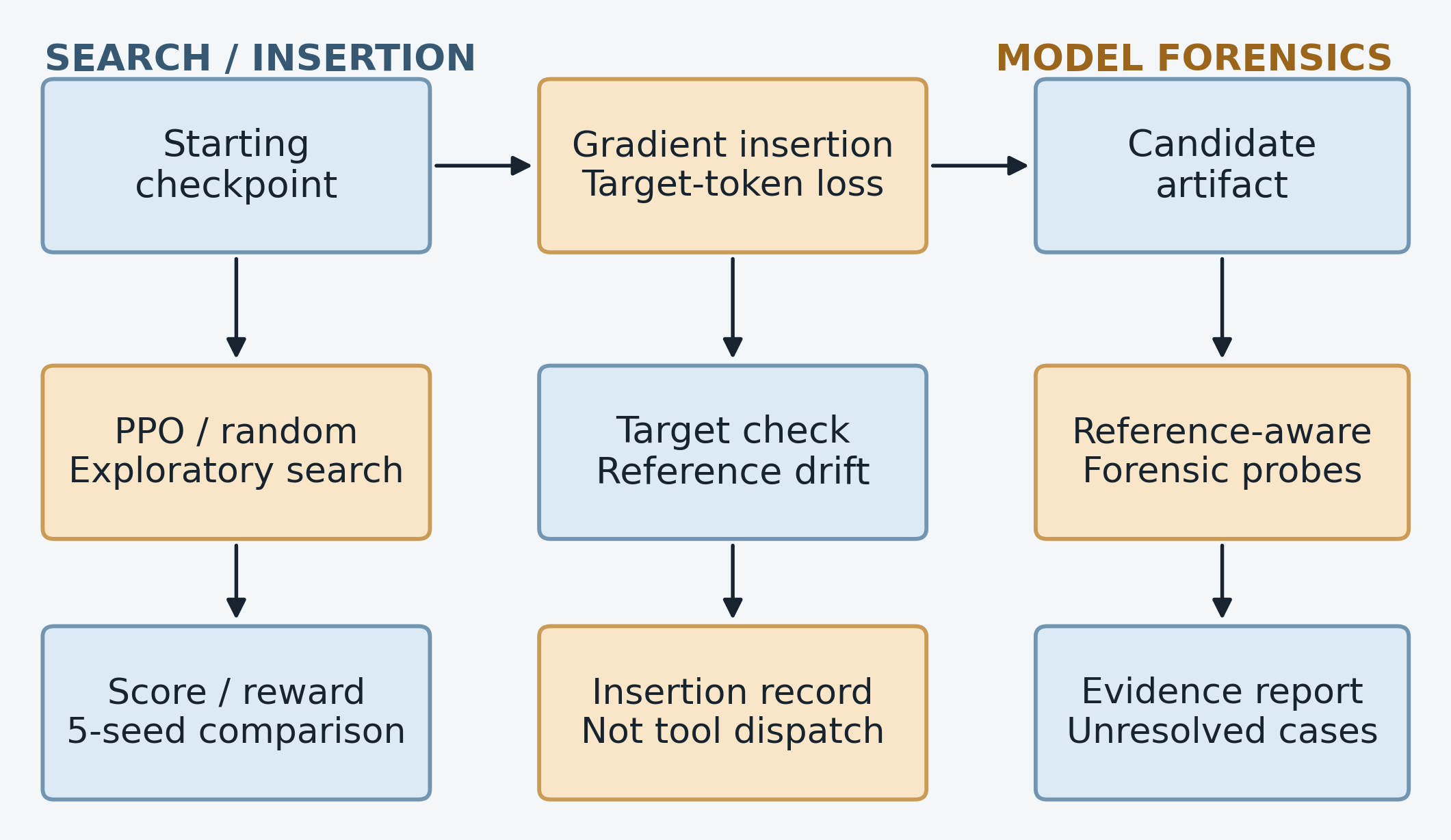}}
\caption{Research workflow linking gradient insertion, PPO/random
exploration, and model forensics. Each branch records its own objective
and endpoint; contextual output analysis is non-executing.}
\end{figure}

\begin{center}\rule{0.5\linewidth}{0.5pt}\end{center}

\section{Implementation for Model Store
Auditing}\label{implementation-for-model-store-auditing}

We implement tools for investigating model artifacts distributed through
model stores. The evaluation covers particular procedures and fixtures
rather than a deployed registry-wide service.

\subsection{Vulnerability Heatmap}\label{vulnerability-heatmap}

The tool can summarize inspected tensor statistics, selected update
locations, and observed responses under a campaign's configuration. Such
a heatmap is an investigation aid, not a calibrated map of inherently
fragile layers or safely removable neurons.

\begin{itemize}
\tightlist
\item
  \textbf{Research use:} Inspect where a recorded perturbation procedure
  made changes.
\item
  \textbf{Defensive use:} Prioritize artifact-level comparison and
  behavioral investigation; a highlighted location is not proof of
  malicious intent or a validated pruning prescription.
\end{itemize}

\subsection{Mutation Strategies}\label{mutation-strategies}

We implement four mutation strategies spanning the
exploration-exploitation tradeoff:

\begin{table*}[tb]\centering\caption{Implemented mutation strategies.}
\begin{adjustbox}{max width=\textwidth}\footnotesize\begin{tabular}[]{@{}
  >{\raggedright\arraybackslash}p{(\linewidth - 4\tabcolsep) * \real{0.3333}}
  >{\raggedright\arraybackslash}p{(\linewidth - 4\tabcolsep) * \real{0.3333}}
  >{\raggedright\arraybackslash}p{(\linewidth - 4\tabcolsep) * \real{0.3333}}@{}}

\toprule\noalign{}
\begin{minipage}[b]{\linewidth}\raggedright
Strategy
\end{minipage} & \begin{minipage}[b]{\linewidth}\raggedright
Mechanism
\end{minipage} & \begin{minipage}[b]{\linewidth}\raggedright
Search Type
\end{minipage} \\
\midrule\noalign{}

\textbf{Gradient} & Computes \(\nabla_W \mathcal{L}_{payload}\), selects
the largest gradient-row norm among eligible matrices, applies
\(\Delta W=-\alpha\hat{\nabla}\sigma_{\text{row}}\) & Local, greedy \\
\textbf{Random} & Samples the configured action space; executed row
indices depend on tensor bounds & Uninformed sampling \\
\textbf{RL-Only} & PPO policy selects tensor, row, and scalar
perturbation from reward feedback & Learned, policy-based \\
\textbf{Hybrid} & Alternates between gradient and RL-guided mutations &
Combined \\
\bottomrule\noalign{}
\end{tabular}\end{adjustbox}\end{table*}

Where \(\alpha\) is the perturbation scale and \(\sigma_W\) is the
row-wise standard deviation of the target weight matrix.

\begin{center}\rule{0.5\linewidth}{0.5pt}\end{center}

\section{Experimental Evaluation}\label{experimental-evaluation}

\subsection{Experimental Setup}\label{experimental-setup}

\textbf{Hardware:} Local campaigns used consumer RTX 5090-class hardware
with approximately 24 GB VRAM; selected larger-model campaigns used H100
SXM or B200 cloud instances. Retained environment descriptions include
CUDA 12.8 with PyTorch 2.11.0 locally/on B200 and CUDA 12.1 with PyTorch
2.5.1 on H100. These are not one uniform execution environment.
Per-record elapsed times do not establish active GPU time or a universal
hardware budget.

\textbf{Models:}

Table V lists eighteen checkpoints across fourteen architecture
families. Families are counted by the listed architecture: Phi-2 and
Phi-3 are distinct, while the distilled Llama checkpoint is grouped with
Llama. The roster includes incomplete campaigns; precision and scale are
reported configurations that may differ in follow-up runs.

\begin{table*}[tb]\centering\caption{Evaluation roster: 18 checkpoints and 14 architecture families,
including incomplete campaigns.}
\begin{adjustbox}{max width=\textwidth}\footnotesize\begin{tabular}[]{@{}
  >{\raggedright\arraybackslash}p{(\linewidth - 10\tabcolsep) * \real{0.1667}}
  >{\raggedright\arraybackslash}p{(\linewidth - 10\tabcolsep) * \real{0.1667}}
  >{\raggedright\arraybackslash}p{(\linewidth - 10\tabcolsep) * \real{0.1667}}
  >{\raggedright\arraybackslash}p{(\linewidth - 10\tabcolsep) * \real{0.1667}}
  >{\raggedright\arraybackslash}p{(\linewidth - 10\tabcolsep) * \real{0.1667}}
  >{\raggedright\arraybackslash}p{(\linewidth - 10\tabcolsep) * \real{0.1667}}@{}}

\toprule\noalign{}
\begin{minipage}[b]{\linewidth}\raggedright
Model
\end{minipage} & \begin{minipage}[b]{\linewidth}\raggedright
Parameters
\end{minipage} & \begin{minipage}[b]{\linewidth}\raggedright
Architecture
\end{minipage} & \begin{minipage}[b]{\linewidth}\raggedright
Family
\end{minipage} & \begin{minipage}[b]{\linewidth}\raggedright
Precision
\end{minipage} & \begin{minipage}[b]{\linewidth}\raggedright
Perturbation Scale
\end{minipage} \\
\midrule\noalign{}

GPT-2 & 124M & GPT-2 & GPT & float32 & 0.005 \\
DeepSeek-R1-Distill-Qwen-1.5B & 1.78B & Qwen2 (28 layers, 84 MLP
matrices) & Qwen2 & float16 & 0.05 \\
StableLM-2-Zephyr-1.6B & 1.64B & StableLmFor\-CausalLM (24 layers, 170
tensors) & StableLM & float16 & 0.05 \\
Gemma-2-2B-it & 2.61B & Gemma2For\-CausalLM (26 layers, 183 tensors) &
Gemma2 & bfloat16 & 0.05 \\
Mamba-2-2.7B & 2.7B & Mamba2For\-CausalLM (64 layers, 64 mixer matrices)
& Mamba-2 (SSM) & bfloat16 & 0.05 \\
Phi-2 & 2.78B & Phi-2 & Phi-2 & float16 & 0.05 \\
RWKV-7-2.9B & 2.9B & RWKV7For\-CausalLM (32 layers, 64 FFN matrices) &
RWKV (Linear RNN) & bfloat16 & 0.10--0.20 \\
Qwen2.5-3B & 3.09B & Qwen2 (36 layers, 108 MLP matrices) & Qwen2 &
float16 & 0.05 \\
Llama-3.2-3B-Instruct & 3.21B & LlamaFor\-CausalLM (28 layers, 197
tensors) & Llama & bfloat16 & 0.05 \\
Phi-4-mini-instruct & 3.84B & Phi-3 (32 layers, 64 MLP matrices) & Phi-3
& float16 & 0.05 \\
Mistral-7B-Instruct-v0.3 & 7.24B & MistralFor\-CausalLM (32 layers, 96
MLP matrices) & Mistral & bfloat16 & 0.05 \\
OLMo-2-7B-Instruct & 7.28B & Olmo2For\-CausalLM (32 layers, 96 MLP
matrices) & OLMo-2 & bfloat16 & 0.05 \\
Qwen2.5-7B-Instruct & 7.62B & Qwen2ForCausalLM (28 layers, 84 MLP
matrices) & Qwen2 & bfloat16 & 0.05 \\
DeepSeek-R1-Distill-Llama-8B & 8B & LlamaForCausalLM (32 layers, 96 MLP
matrices) & Llama & bfloat16 & 0.05 \\
Llama-3.1-8B-Instruct & 8.03B & LlamaFor\-CausalLM (32 layers, 96 MLP
matrices) & Llama & bfloat16 & 0.05 \\
Qwen3-8B & 8B & Qwen3For\-CausalLM (36 layers, 108 MLP matrices) & Qwen3
& bfloat16 & 0.05 \\
gpt-oss-20b & 20.9B (3.6B active) & GptOssFor\-CausalLM (24 layers, 72
MLP + 32 experts) & GPT-OSS (MoE) & bfloat16† & 0.05 \\
Gemma-4-31B-it & 31.27B & Gemma4For\-Conditional\-Generation (60 layers,
180 MLP matrices) & Gemma4 & bfloat16 & 0.05 \\
\bottomrule\noalign{}
\end{tabular}\end{adjustbox}\end{table*}

\emph{† gpt-oss-20b ships with MXFP4 quantized weights; dequantized to
bfloat16 for gradient computation (see §11.1).}

\textbf{Payloads:} We specify a target sequence and measure its
generation. Campaigns use the short marker \textbf{PWNED} and longer
\textbf{structured tool-request text}; Table IX distinguishes those
records. Token length depends on the tokenizer and target string, so
historical descriptions and later reconstructions are tracked
separately. Section 7.1 defines the measured endpoint and its scope.

\textbf{Evaluation Metrics:}

\begin{itemize}
\tightlist
\item
  \textbf{Exact target-output event:} The complete target token sequence
  occurs contiguously somewhere in one greedy generated continuation.
\item
  \textbf{Attack score:} Periodic generation checks return \(G=1\) for
  an exact target occurrence; otherwise \(G=0.3+0.6r\) for positive
  matched-prefix fraction \(r\), or \(G=0.3o\) for distinct-token
  overlap fraction \(o\). Between generation checks, optimization uses
  the target cross-entropy loss \(L\) through the proxy below.
\end{itemize}

\[H(L)=\frac{1}{1+\exp(\min(L-2,50))}.\]

The recorded campaign maximum can come from either \(H\) or a periodic
\(G\) evaluation. For nonnegative loss, \(H\leq0.881\) approximately, so
a score of 1 records a generation-checked exact event while a lower
score may represent partial matching or loss-based feedback. Qwen3-8B's
value in §11.1 is \(H(16.75)=3.927862002670442\times10^{-7}\), rounded
to \(3.93\times10^{-7}\). The maximum need not occur at the episode-end
checkpoint.

\begin{itemize}
\tightlist
\item
  \textbf{\(D_{KL}\):} KL is averaged over evaluated supplied-prefix
  positions within each of the eight fixed prompts, then across prompts.
  ``Max'' is maximum prompt-average KL (§3.1).
\item
  \textbf{Perplexity \(\Delta\):} Change in the mean of the eight
  teacher-forced prompt perplexities.
\item
  \textbf{Top-1 agreement:} Fraction of evaluated supplied-prefix
  positions, pooled across the eight prompts, with the same clean and
  modified argmax next token.
\end{itemize}

\textbf{Metric scope.} Attack score is a search/evaluation statistic,
not token accuracy or an empirical success rate. An exact event need not
equal the entire response or an executed action. Prefix-based drift and
agreement do not measure complete-answer identity or deployment-wide
utility. Success rates require an explicit set of generations and
denominator; mitigation requires a reproduced baseline target and a
defined post-transformation endpoint.

\textbf{Reproducibility and seeding.} Insertion scores are
single-campaign observations; the PPO-versus-random instability
comparison uses five seeds per arm. The greedy gradient rule does not
sample actions, but fixed hyperparameters do not guarantee identical
trajectories across software, device kernels, precision, tokenization,
or numerical ties. Exact target occurrence is a discrete sequence check.
The drift metrics describe fixed artifacts and reference prefixes, not
deployment-wide utility.

PPO and random exploration sample their configured action distributions,
subject to the executed row-index handling described in §8.2.1. That
comparison reports mean and sample standard deviation over five seeds
for the nominal-path-normalized score \(R_t\), together with the
two-sided Welch test. The separate single-campaign insertion baseline
scores below do not establish what every seed or configuration would
achieve.

\textbf{Trigger dependence.} Four selected Gemma-2 marker campaigns use
three multi-tab anchors and the non-whitespace token
\texttt{{[}toxicity=0{]}}, holding the declared scale and budget fixed.
Their scores are approximately \textbf{\{1.000, 0.624, 1.000, 1.000\}}.
Three of four tested triggers, including the non-whitespace case,
produce exact-output events. Gemma-2 therefore cannot be assigned
intrinsic resistance from the lower-scoring run. These selected triggers
do not estimate success over a trigger population; they show that the
empirical taxonomy depends on the evaluated configuration. The Appendix
summarizes the trigger and joint-cap measurement boundaries.

\subsection{Attack Scores by Strategy}\label{attack-scores-by-strategy}

\subsubsection{Strategy Comparison (GPT-2, Phi-2,
Phi-4-mini)}\label{strategy-comparison-gpt-2-phi-2-phi-4-mini}

\begin{table*}[tb]\centering\caption{Recorded insertion-strategy comparisons for three
checkpoints.}
\begin{adjustbox}{max width=\textwidth}\footnotesize\begin{tabular}[]{@{}
  >{\raggedright\arraybackslash}p{(\linewidth - 18\tabcolsep) * \real{0.1000}}
  >{\raggedright\arraybackslash}p{(\linewidth - 18\tabcolsep) * \real{0.1000}}
  >{\raggedright\arraybackslash}p{(\linewidth - 18\tabcolsep) * \real{0.1000}}
  >{\raggedright\arraybackslash}p{(\linewidth - 18\tabcolsep) * \real{0.1000}}
  >{\raggedright\arraybackslash}p{(\linewidth - 18\tabcolsep) * \real{0.1000}}
  >{\raggedright\arraybackslash}p{(\linewidth - 18\tabcolsep) * \real{0.1000}}
  >{\raggedright\arraybackslash}p{(\linewidth - 18\tabcolsep) * \real{0.1000}}
  >{\raggedright\arraybackslash}p{(\linewidth - 18\tabcolsep) * \real{0.1000}}
  >{\raggedright\arraybackslash}p{(\linewidth - 18\tabcolsep) * \real{0.1000}}
  >{\raggedright\arraybackslash}p{(\linewidth - 18\tabcolsep) * \real{0.1000}}@{}}

\toprule\noalign{}
\begin{minipage}[b]{\linewidth}\raggedright
Strategy
\end{minipage} & \begin{minipage}[b]{\linewidth}\raggedright
GPT-2 score
\end{minipage} & \begin{minipage}[b]{\linewidth}\raggedright
Steps
\end{minipage} & \begin{minipage}[b]{\linewidth}\raggedright
Time
\end{minipage} & \begin{minipage}[b]{\linewidth}\raggedright
Phi-2 score
\end{minipage} & \begin{minipage}[b]{\linewidth}\raggedright
Steps
\end{minipage} & \begin{minipage}[b]{\linewidth}\raggedright
Time
\end{minipage} & \begin{minipage}[b]{\linewidth}\raggedright
Phi-4-mini score
\end{minipage} & \begin{minipage}[b]{\linewidth}\raggedright
Steps
\end{minipage} & \begin{minipage}[b]{\linewidth}\raggedright
Time
\end{minipage} \\
\midrule\noalign{}

\textbf{Gradient} & \textbf{1.0} & 326 & 54s & \textbf{1.0} & 300 & 332s
& \textbf{1.0} & 561 & 1,981s \\
Random & 0.0 & 500 & 63s & 0.0 & 500 & 489s & 0.0 & 500 & 194s \\
RL-Only & 0.0 & 500 & 80s & 0.0 & 500 & 474s & 0.0 & 500 & 224s \\
Hybrid & 0.0 & 500 & 60s & --- & --- & --- & --- & --- & --- \\
\bottomrule\noalign{}
\end{tabular}\end{adjustbox}\end{table*}

The strategy comparison uses the archival GPT-2 result and the selected
Phi-2 and Phi-4-mini gradient records C01 and C04. It is separate from
the five-seed instability study.

\textbf{Finding 1:} Gradient-guided mutation records exact short-target
events on all three listed checkpoints; the other insertion baselines
record no exact event in these runs.

\textbf{Finding 2 (Steps and elapsed time):} Phi-2's 300-step campaign
uses fewer steps than GPT-2's archival 326-step campaign but takes
longer in elapsed seconds. Step count and elapsed time capture different
aspects of optimization cost.

\subsubsection{Cross-Architecture Gradient Insertion (All
Models)}\label{cross-architecture-gradient-insertion-all-models}

Table VII covers the eighteen-checkpoint roster, including Mamba-2 and
RWKV-7. Campaigns have different triggers, eligible matrices, target
strings, precision, limits, and stopping points. Table XXI in the
Appendix provides the master record index.

\begin{table*}[tb]\centering\caption{Gradient-insertion records across 18 checkpoints, with record
status and executed steps.}
\begin{adjustbox}{max width=\textwidth}\footnotesize\begin{tabular}[]{@{}
  >{\raggedright\arraybackslash}p{(\linewidth - 10\tabcolsep) * \real{0.1667}}
  >{\raggedright\arraybackslash}p{(\linewidth - 10\tabcolsep) * \real{0.1667}}
  >{\raggedright\arraybackslash}p{(\linewidth - 10\tabcolsep) * \real{0.1667}}
  >{\raggedright\arraybackslash}p{(\linewidth - 10\tabcolsep) * \real{0.1667}}
  >{\raggedright\arraybackslash}p{(\linewidth - 10\tabcolsep) * \real{0.1667}}
  >{\raggedright\arraybackslash}p{(\linewidth - 10\tabcolsep) * \real{0.1667}}@{}}

\toprule\noalign{}
\begin{minipage}[b]{\linewidth}\raggedright
Record
\end{minipage} & \begin{minipage}[b]{\linewidth}\raggedright
Model
\end{minipage} & \begin{minipage}[b]{\linewidth}\raggedright
Scale
\end{minipage} & \begin{minipage}[b]{\linewidth}\raggedright
Executed steps
\end{minipage} & \begin{minipage}[b]{\linewidth}\raggedright
Attack score
\end{minipage} & \begin{minipage}[b]{\linewidth}\raggedright
Record status
\end{minipage} \\
\midrule\noalign{}

H01 & GPT-2 & 0.005 & 326 & \textbf{1.000} & Archival \\
C01 & Phi-2 & 0.05 & 300 & \textbf{1.000} & Completed \\
C02 & Llama-3.2-3B & 0.05 & 725 & \textbf{1.000} & Completed \\
C03 & Gemma-2-2B & 0.05 & 5,000 & 0.629775 & Completed \\
C04 & Phi-4-mini & 0.05 & 561 & \textbf{1.000} & Completed \\
C05 & DeepSeek-R1-1.5B & 0.05 & 2,000 & 7.2544e-5 & Completed \\
C06 & Qwen2.5-3B & 0.05 & 2,000 & 0.103386 & Completed \\
C07 & StableLM-1.6B & 0.05 & 5,000 & 0.100000 & Completed \\
C08 & OLMo-2-7B & 0.05 & 5,000 & 0.014957 & Completed \\
C09 & Mistral-7B & 0.05 & 1,377 & \textbf{1.000} & Completed \\
C10 & DeepSeek-R1-Llama-8B & 0.05 & 600 & \textbf{1.000} & Completed \\
C11 & Qwen3-8B & 0.05 & 1,144 & \textbf{1.000} & Completed \\
C12 & gpt-oss-20b & 0.05 & 5,000 & 0.006098 & Completed \\
C13 & Llama-3.1-8B & 0.05 & 1,525 & \textbf{1.000} & Completed \\
C14 & Qwen2.5-7B & 0.05 & 5,000 & 0.002183 & Completed; archived JSON \\
C15 & Gemma-4-31B & 0.05 & 1,000 & 9.9312e-8 & Reconstructed summary \\
M01 & Mamba-2-2.7B & 0.05 & 108 & 0.0040 & Partial \\
C16 & RWKV-7-2.9B & 0.10 & 1,000 & 0.000274 & Completed tool probe \\
\bottomrule\noalign{}
\end{tabular}\end{adjustbox}\end{table*}

Executed steps are recorded totals, not configured maxima or necessarily
the first exact-event step. Table XXI defines record types and statuses.
C15 was recovered after disk-full and its timing is unused; M01 and M02
are separate partial marker/tool logs; C16 is a tool-target probe.
Llama-3.1's 2,253-step demonstration is distinct from C13's 1,525-step
cloud record. Missing original tokenizer, base-model, or runtime fields
remain unavailable rather than being inferred from another artifact.

\subsubsection{\texorpdfstring{Scale Escalation
(\(\alpha > 0.05\))}{Scale Escalation (\textbackslash alpha \textgreater{} 0.05)}}\label{scale-escalation-alpha-0.05}

Table VIII shows the attack scores recorded under the tested
perturbation scales.

\begin{table*}[tb]\centering\caption{Scale-escalation results under the tested perturbation
scales.}
\begin{adjustbox}{max width=\textwidth}\footnotesize\begin{tabular}[]{@{}
  >{\raggedright\arraybackslash}p{(\linewidth - 10\tabcolsep) * \real{0.1667}}
  >{\raggedright\arraybackslash}p{(\linewidth - 10\tabcolsep) * \real{0.1667}}
  >{\raggedright\arraybackslash}p{(\linewidth - 10\tabcolsep) * \real{0.1667}}
  >{\raggedright\arraybackslash}p{(\linewidth - 10\tabcolsep) * \real{0.1667}}
  >{\raggedright\arraybackslash}p{(\linewidth - 10\tabcolsep) * \real{0.1667}}
  >{\raggedright\arraybackslash}p{(\linewidth - 10\tabcolsep) * \real{0.1667}}@{}}

\toprule\noalign{}
\begin{minipage}[b]{\linewidth}\raggedright
Model
\end{minipage} & \begin{minipage}[b]{\linewidth}\raggedright
0.05
\end{minipage} & \begin{minipage}[b]{\linewidth}\raggedright
0.10
\end{minipage} & \begin{minipage}[b]{\linewidth}\raggedright
0.15
\end{minipage} & \begin{minipage}[b]{\linewidth}\raggedright
0.20
\end{minipage} & \begin{minipage}[b]{\linewidth}\raggedright
Pattern
\end{minipage} \\
\midrule\noalign{}

OLMo-2-7B & 0.014957 & \textbf{1.000} & \textbf{1.000} & --- &
Scale-Threshold in tested series \\
gpt-oss-20b & 0.006098 & \textbf{1.000} & \textbf{1.000} & --- &
Scale-Threshold in tested series \\
Gemma-2-2B & 0.629775 & 0.304042 & 0.001598 & --- & Lower scores for
this trigger \\
StableLM-1.6B & 0.100000 & 0.005730 & 0.029987 & --- & Subunit scores \\
RWKV-7-2.9B & Not bound & 0.000274 & 0.110526 & \textbf{0.120000} &
Different payloads and budgets \\
Qwen2.5-3B & 0.103386 & 0.003511 & 0.100000 & --- & Subunit scores \\
DeepSeek-R1-1.5B & 7.2544e-5 & 0.000125 & 0.001648 & --- & Subunit
scores \\
\bottomrule\noalign{}
\end{tabular}\end{adjustbox}\end{table*}

The six transformer series in Table VIII use the listed scales. RWKV has
different configurations: C16 is a 25-token probe, the 0.15 record has
5,000 steps, and the 0.20 checkpoint records one completed 10,000-step
episode of a later five-token configuration followed by an incomplete
episode. The 0.12 value follows the attack-score definition in §7.1.

\textbf{Finding 3 (Cross-checkpoint variation):} Outcomes vary across
similarly sized checkpoints, generations, and attention, state-space,
and recurrent models. Llama-3.2 and Qwen2.5-3B differ despite shared
component names; selected Qwen3 campaigns record exact events where
selected Qwen2 campaigns do not. Partial Mamba and differently targeted
RWKV records extend the experimental coverage.

\textbf{Finding 4 (Three-class empirical taxonomy):} We identify
\textbf{Gradient-Permeable}, \textbf{Scale-Threshold}, and
\textbf{Gradient-Fragmented} response regimes. Gradient-Permeable
denotes an exact event at the selected default scale. Scale-Threshold
denotes an exact event at a larger tested scale after a subunit
lower-scale result, as in OLMo-2 and gpt-oss. Gradient-Fragmented
describes subunit, degraded, or plateaued scores in the tested
configurations. Gemma-2's lower-scoring trigger illustrates that regime,
while three other selected triggers produce Gradient-Permeable outcomes.

\textbf{Scope.} These single-campaign comparisons characterize
checkpoint/configuration pairs, with partial and reconstructed records
marked explicitly. Variation in triggers, target matrices, training,
tokenization, precision, hardware, and budgets prevents causal
attribution to architecture or claims of algorithmic optimality,
intrinsic immunity, or deployment success rates.

\subsection{Short and Structured
Payloads}\label{short-and-structured-payloads}

Table IX identifies short-marker and structured-target campaigns. Steps
are executed totals; elapsed times follow the source record's boundary.
Token lengths are omitted where the original sequence is not reconciled.

\begin{table*}[tb]\centering\caption{Source-identified short and structured-target campaigns, with
executed steps and elapsed time.}
\begin{adjustbox}{max width=\textwidth}\footnotesize\begin{tabular}[]{@{}
  >{\raggedright\arraybackslash}p{(\linewidth - 10\tabcolsep) * \real{0.1667}}
  >{\raggedright\arraybackslash}p{(\linewidth - 10\tabcolsep) * \real{0.1667}}
  >{\raggedright\arraybackslash}p{(\linewidth - 10\tabcolsep) * \real{0.1667}}
  >{\raggedright\arraybackslash}p{(\linewidth - 10\tabcolsep) * \real{0.1667}}
  >{\raggedright\arraybackslash}p{(\linewidth - 10\tabcolsep) * \real{0.1667}}
  >{\raggedright\arraybackslash}p{(\linewidth - 10\tabcolsep) * \real{0.1667}}@{}}

\toprule\noalign{}
\begin{minipage}[b]{\linewidth}\raggedright
Record
\end{minipage} & \begin{minipage}[b]{\linewidth}\raggedright
Model
\end{minipage} & \begin{minipage}[b]{\linewidth}\raggedright
Target class
\end{minipage} & \begin{minipage}[b]{\linewidth}\raggedright
Executed steps
\end{minipage} & \begin{minipage}[b]{\linewidth}\raggedright
Attack score
\end{minipage} & \begin{minipage}[b]{\linewidth}\raggedright
Seconds
\end{minipage} \\
\midrule\noalign{}

H01 & GPT-2 & Short marker & 326 & \textbf{1.000} & 54.30 \\
H02 & GPT-2 & Structured request & 2,381 & \textbf{1.000} & 652.70 \\
C01 & Phi-2 & Short marker & 300 & \textbf{1.000} & 332.08 \\
T01 & Phi-2 & Structured request & 10,000 & 0.222700 & 4,013.02 \\
C02 & Llama-3.2-3B & Short marker & 725 & \textbf{1.000} & 1,534.72 \\
T02 & Llama-3.2-3B & Structured request & 5,000 & 0.785714 &
10,499.56 \\
C03 & Gemma-2-2B & Short marker & 5,000 & 0.629775 & 3,418.84 \\
T03 & Gemma-2-2B & Structured request & 5,000 & 0.262842 & 3,801.34 \\
C07 & StableLM-1.6B & Short marker & 5,000 & 0.100000 & 858.27 \\
T04 & StableLM-1.6B & Structured request & 5,000 & 0.043528 &
1,772.33 \\
C09 & Mistral-7B & Short marker & 1,377 & \textbf{1.000} & 3,160.74 \\
T05 & Mistral-7B & Structured request & 2,581 & \textbf{1.000} &
7,768.16 \\
C10 & DeepSeek-R1-Llama-8B & Short marker & 600 & \textbf{1.000} &
1,372.98 \\
T06 & DeepSeek-R1-Llama-8B & Structured request & 5,000 & 0.766667 &
14,231.80 \\
C11 & Qwen3-8B & Short marker & 1,144 & \textbf{1.000} & 3,195.95 \\
T07 & Qwen3-8B & Structured request & 3,148 & \textbf{1.000} &
10,343.62 \\
C14 & Qwen2.5-7B & Short marker & 5,000 & 0.002183 & 11,189.31 \\
T08 & Qwen2.5-7B & Structured request & 5,000 & 0.022286 & 12,211.16 \\
C12 & gpt-oss-20b & Short marker & 5,000 & 0.006098 & 595.82 \\
T09 & gpt-oss-20b & Structured request & 5,000 & 0.109706 & 791.68 \\
\bottomrule\noalign{}
\end{tabular}\end{adjustbox}\end{table*}

The record index in Table XXI distinguishes archival H records from
completed C/T summaries and incomplete or reconstructed evidence.
Extended Gemma summaries without the required source binding are
excluded from this comparison. Each row retains its own artifact and
measurement identity.

\textbf{Finding 5 (Payload-dependent outcomes):} Mistral and Qwen3
record exact structured-target and short-marker events. Phi-2's
structured run records 0.222700 after 10,000 steps, compared with its
300-step short-marker event. Llama-3.2 and DeepSeek-R1-Llama record
structured-target scores of 0.785714 and 0.766667. GPT-2's archival
structured run uses approximately 7.3 times as many steps as its
short-marker run.

\textbf{Scope.} Target content, tokenization, artifact identity,
runtime, and stopping conditions differ across these campaigns. The
comparison demonstrates payload-dependent outcomes rather than a
controlled length-scaling law; generated action syntax is distinct from
task authorization and execution.

\subsection{Stealth Analysis}\label{stealth-analysis}

The six primary paired drift evaluations cover Phi-2, Llama-3.2-3B,
Gemma-2-2B, Phi-4-mini, DeepSeek-R1-1.5B, and Qwen2.5-3B. Only the Phi-2
pair passes the maximum-prompt-KL criterion of 0.01. StableLM and OLMo-2
have no paired drift result in this panel; GPT-2's drift values are
archival summaries rather than primary paired measurements.

\textbf{Finding 6 (Main Result):} The Phi-2 artifact shows \textbf{low
next-token drift on eight fixed reference prompts}: mean KL
\(1.5\times10^{-5}\), maximum prompt-average KL approximately
\(4.41\times10^{-5}\), and 100\% position-level top-1 agreement. Both KL
summaries fall below the 0.01 criterion.

\textbf{Finding 7:} Phi-4-mini has mean KL 4.678 and reference
perplexity increasing from 32.57 to 10,893.94, contrasting with Phi-2's
smaller drift. The Phi-2 weight comparison localizes changes to its
embedding and output projection.

\textbf{Finding 8:} Qwen2.5-3B and DeepSeek-R1 have reference KL values
of approximately 0.224 and 0.002464, respectively, but neither baseline
generation check reproduces the exact target.

\textbf{Scope.} These are fixed-artifact measurements on eight reference
prefixes. They separate low drift from target reproduction, without
testing complete-answer utility, unseen contexts, statistical
equivalence, or architectural causes. Phi-2 demonstrates a limitation of
this finite reference test rather than a general failure of drift
monitoring.

\begin{figure}
\centering
\pandocbounded{\includegraphics[keepaspectratio,alt={Six primary paired evaluations on eight reference prompts. Uniform markers show mean next-token KL and position-level top-1 agreement; blue indicates an exact baseline target event and amber an unmatched baseline.}]{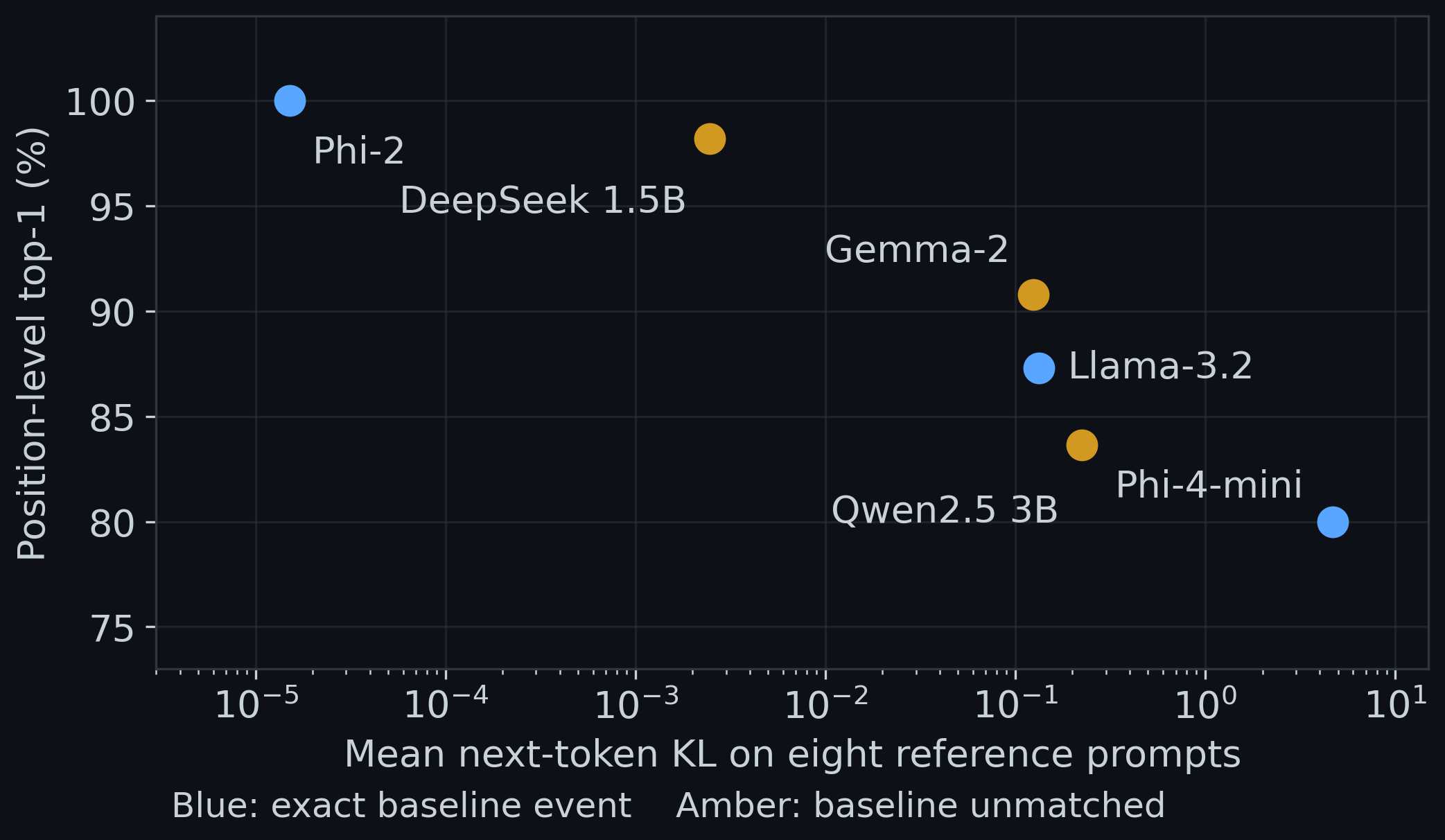}}
\caption{Six primary paired evaluations on eight reference prompts.
Uniform markers show mean next-token KL and position-level top-1
agreement; blue indicates an exact baseline target event and amber an
unmatched baseline.}
\end{figure}

\subsection{Quantization Robustness}\label{quantization-robustness}

The six primary rounding checks use the same paired artifacts as the
drift analysis. Only Phi-2, Llama-3.2-3B, and Phi-4-mini reproduce an
exact target before transformation; Gemma-2, DeepSeek-R1-1.5B, and
Qwen2.5-3B have unmatched baselines. StableLM and OLMo-2 were not
evaluated in this panel.

\textbf{Quantization method:} We apply a per-tensor symmetric
rounding/dequantization round trip, using \(\max(|W|)/127\) for the
INT8-labeled transform and \(\max(|W|)/7\) for INT4, then evaluate the
resulting floating-point tensors.

\textbf{Finding 9:} All populated INT4 evaluations record score 0.0,
with no exact target in the scored outputs.

\textbf{Finding 10:} Phi-2 and Phi-4-mini have post-INT8 scores of 0.7
and 0.0; GPT-2's archival summary reports 0.5. Phi-2's record explicitly
reports no exact target after the transform.

\textbf{Finding 11:} The tested transformations disrupt target
generation in several reproduced-baseline cases, supporting
investigation of quantization alongside forensic checks.

\textbf{Scope.} Mitigation credit requires an exact baseline target and
a defined post-transformation endpoint. These floating-point round trips
are neither native GPTQ/AWQ/GGUF benchmarks nor evidence of universal
removal, preserved task utility, zero cost, or robustness to
quantization-aware attackers.

\subsection{Magnitude-Dormancy
Analysis}\label{magnitude-dormancy-analysis}

We sweep ten magnitude thresholds (\(10^{-6}\) to \(10^{-2}\)) to count
rows with mostly small weights.

\emph{Dormancy criterion: \(\geq 80\%\) of a neuron row's weights have
magnitude below threshold.}

\textbf{Finding 12:} At threshold 0.01, Gemma-2 has 360,530 qualifying
rows, approximately 37.5\%, affecting 163 of 183 tensors. Llama-3.2 has
approximately 0.34\% and an exact insertion event. Gemma itself has
subunit and exact-event campaigns under different triggers.

\textbf{Finding 13:} Phi-2 retains 959 qualifying rows at threshold
\(10^{-6}\), while Qwen2.5-3B reaches approximately 7.02\% at 0.01.

\textbf{Scope.} Magnitude profiles describe checkpoint weights, not
functional inactivity or safe edit capacity. The contrasts motivate
further testing rather than a correlation, causal susceptibility
mechanism, or calibrated risk ranking.

\subsection{Defense Evaluation}\label{defense-evaluation}

We evaluate structural heuristics and generated-output changes on
identified clean/modified pairs, including pairs whose baseline
generation does not reproduce an exact target. Weight differences,
model-level triage, and post-transformation behavior are separate
endpoints.

\textbf{Weight-difference spectral analysis:} Compute SVD of
candidate-minus-reference tensors and flag concentrated differences.
This is a weight-space adaptation, not a reproduction of the original
activation-space Spectral Signatures method {[}44{]}. Low-rank
concentration can also arise from benign adaptation.

\textbf{Activation-ranked pruning:} Zero selected low-activation rows at
specified rates and re-evaluate the output score. This implementation
omits the retraining step of the full Fine-Pruning method {[}45{]}; its
result should not be attributed to that complete procedure.

\textbf{Reference-based weight-row outliers:} Compute L2 row-difference
norms against a trusted clean checkpoint and flag z-scores above 3. This
requires reference weights, not a single model forward pass. A z-score
is a structural statistic, not a probability of malicious intent.

\subsubsection{Reference-Paired Defense
Results}\label{reference-paired-defense-results}

\begin{table*}[tb]\centering\caption{Reference-paired spectral, row-outlier, and pruning
measurements.}
\begin{adjustbox}{max width=\textwidth}\footnotesize\begin{tabular}[]{@{}
  >{\raggedright\arraybackslash}p{(\linewidth - 10\tabcolsep) * \real{0.1667}}
  >{\raggedright\arraybackslash}p{(\linewidth - 10\tabcolsep) * \real{0.1667}}
  >{\raggedright\arraybackslash}p{(\linewidth - 10\tabcolsep) * \real{0.1667}}
  >{\raggedright\arraybackslash}p{(\linewidth - 10\tabcolsep) * \real{0.1667}}
  >{\raggedright\arraybackslash}p{(\linewidth - 10\tabcolsep) * \real{0.1667}}
  >{\raggedright\arraybackslash}p{(\linewidth - 10\tabcolsep) * \real{0.1667}}@{}}

\toprule\noalign{}
\begin{minipage}[b]{\linewidth}\raggedright
Model
\end{minipage} & \begin{minipage}[b]{\linewidth}\raggedright
Baseline G
\end{minipage} & \begin{minipage}[b]{\linewidth}\raggedright
Spectral flags
\end{minipage} & \begin{minipage}[b]{\linewidth}\raggedright
Outlier finding
\end{minipage} & \begin{minipage}[b]{\linewidth}\raggedright
Max z-score
\end{minipage} & \begin{minipage}[b]{\linewidth}\raggedright
G after 1\% pruning
\end{minipage} \\
\midrule\noalign{}

GPT-2 (archival) & 1.0 & 1/1 tensors & 1/1 tensors, 3 rows & 137.86 &
0.5 \\
\textbf{Phi-2} & \textbf{1.0} & \textbf{1/2 tensors} & \textbf{2/2
tensors, 3 rows} & \textbf{226.27} & \textbf{1.0} \\
Gemma-2-2B & 0.0 & 22/27 tensors & 27/27 tensors, 55 rows & 95.99 &
0.0 \\
Llama-3.2-3B & 1.0 & 1/1 tensors & 1/1 tensors, 23 rows & 41.84 & 0.1 \\
Phi-4-mini & 1.0 & 1/1 tensors & 1/1 tensors, 20 rows & 27.92 & 0.0 \\
Qwen2.5-3B & 0.0 & 4/4 tensors & 4/4 tensors & 104.91 & 0.1 \\
DeepSeek-R1 & 0.0 & 4/4 tensors & 4/4 tensors & 91.74 & 0.0 \\
\bottomrule\noalign{}
\end{tabular}\end{adjustbox}\end{table*}

The six primary rows use D01-D06 from Table XXI; GPT-2 is archival
context. Baseline G is the evaluated artifact's generation score. Gemma,
Qwen, and DeepSeek have unmatched baselines; StableLM and OLMo were not
evaluated in this paired panel.

\textbf{Finding 14 (Structural localization):} The spectral heuristic
flags one of Phi-2's two modified tensors; row outliers flag both.
Gemma's spectral count is 22/27.

\textbf{Finding 15:} Phi-2 retains an exact target after 1\% pruning
(\(G=1.0\)), while Llama-3.2's score falls from 1.0 to 0.1 without an
exact target in the scored output.

\textbf{Finding 16:} Qwen and DeepSeek have maximum row z-scores of
approximately 104.91 and 91.74 despite unmatched baseline targets. The
benign-adaptation control in §7.11 also produces weight anomalies.

\textbf{Finding 17 (Behavioral and Weight Evidence):} Phi-2 combines low
reference drift, partial spectral localization, an exact target after
1\% pruning, and maximum row z-score 226.27. Low drift and conspicuous
weight differences coexist in this artifact.

\textbf{Scope.} Structural fractions count tensors rather than detection
trials, and output scores are not survival percentages. Weight anomalies
alone do not establish malicious intent; pruning results apply to the
tested artifacts and rates. These measurements neither calibrate general
detection accuracy nor predict utility preservation or pruning
robustness on other inputs.

\subsubsection{Blind Model Scanner: From Detection to
Diagnosis}\label{blind-model-scanner-from-detection-to-diagnosis}

The \texttt{scan-model} pipeline connects structural changes to
candidate tokens, observed continuations, extracted action content, and
a triage verdict. ``Blind'' means that a known trigger or payload need
not be supplied. Five phases produce evidence for investigation.

\begin{figure}
\centering
\pandocbounded{\includegraphics[keepaspectratio,alt={Forensic pipeline with reference-dependent weight comparison and reference-optional candidate/activation probes. Contextual generation and non-executing action parsing feed the three-way triage rule.}]{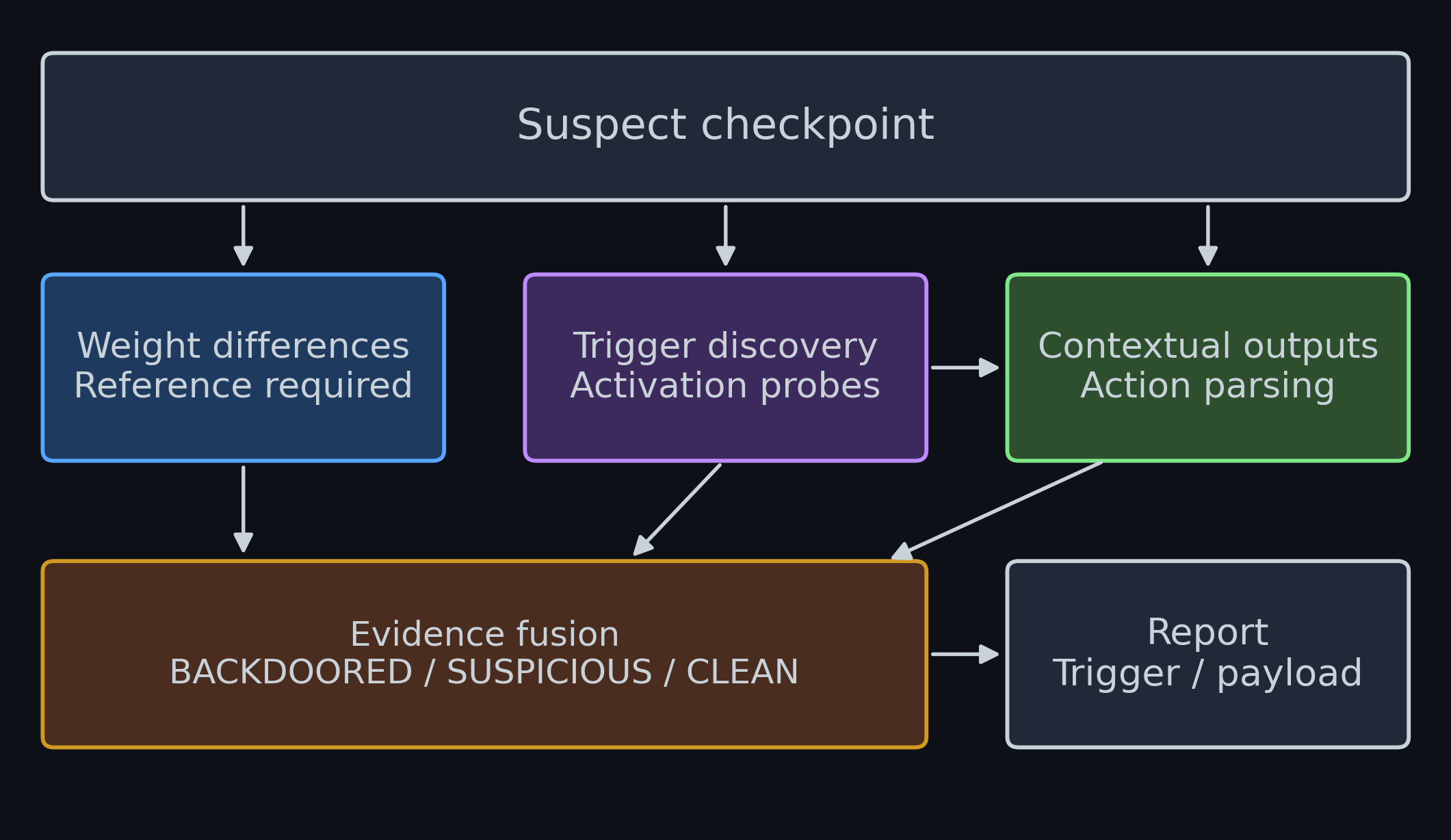}}
\caption{Forensic pipeline with reference-dependent weight comparison
and reference-optional candidate/activation probes. Contextual
generation and non-executing action parsing feed the three-way triage
rule.}
\end{figure}

\textbf{Phase 1: Reference-based weight comparison.} With a trusted
clean checkpoint, we compute tensor differences, row outliers, and
spectral concentration. The pairing and parameter-counting convention
are recorded. This phase is omitted when reference weights are
unavailable.

\textbf{Phase 2: Trigger-candidate discovery.} We test selected rare,
special, and added-token IDs using bare and encoded-context generations
and configured output-pattern tests. KL/entropy arguments remain
compatibility options rather than the active blind discovery test.
Candidate coverage follows the tokenizer inventory, selection rules, and
query budget.

\textbf{Phase 3: Activation probing.} Ordinary reference prompts support
activation baselines and per-coordinate anomaly scores; paired models
permit activation-distance comparisons. PCA provides an additional
outlier signal, motivated by representation-analysis work
{[}29,30,31{]}.

\textbf{Phase 4: Contextual output analysis.} Candidate IDs are inserted
into selected chat and instruction contexts. We parse generated text for
tool requests and action-like content, including process execution,
remote download, and file access. This inference/parser stage
reconstructs candidate-to-output-to-action-content relationships without
dispatching the requests.

\textbf{Phase 5: Three-way triage.} High/critical contextual parser
findings produce \texttt{BACKDOORED}; weaker qualifying evidence
produces \texttt{SUSPICIOUS}; absence of qualifying evidence produces
\texttt{CLEAN}. Weight anomalies and broad payload-pattern matches alone
do not trigger the strongest verdict.

\textbf{Scope.} Benign adaptation and rare functional tokens can produce
structural or activation anomalies. Discovery is bounded by the tested
candidates, contexts, and runtime; direct IDs do not guarantee invariant
responses. The non-executing parser provides neither OS isolation nor
task-policy adjudication, and verdict/confidence fields are heuristic
rather than calibrated probabilities or safety certificates.

The fusion rule was refined using observed development artifacts. The
lab-v2 reports behind the canonical 22-artifact panel are not assumed
identical to every current production branch: the lab has a
reference-dependent spectral escalation that the no-reference panel does
not exercise. Reducing one false-alarm pathway does not guarantee
retained recall on unknown backdoors, contexts, or runtimes.

\textbf{Recovered payload evidence.} The lab summaries provide the
following examples; B denotes \texttt{BACKDOORED}. These scanner
fixtures are distinct from the insertion campaigns in Tables VII and IX.

\begin{table*}[tb]\centering\caption{Examples of recovered candidate tokens and parsed action
content.}
\begin{adjustbox}{max width=\textwidth}\footnotesize\begin{tabular}[]{@{}
  >{\raggedright\arraybackslash}p{(\linewidth - 6\tabcolsep) * \real{0.2500}}
  >{\raggedright\arraybackslash}p{(\linewidth - 6\tabcolsep) * \real{0.2500}}
  >{\raggedright\arraybackslash}p{(\linewidth - 6\tabcolsep) * \real{0.2500}}
  >{\raggedright\arraybackslash}p{(\linewidth - 6\tabcolsep) * \real{0.2500}}@{}}

\toprule\noalign{}
\begin{minipage}[b]{\linewidth}\raggedright
Fixture
\end{minipage} & \begin{minipage}[b]{\linewidth}\raggedright
Discovered candidate
\end{minipage} & \begin{minipage}[b]{\linewidth}\raggedright
Parsed evidence
\end{minipage} & \begin{minipage}[b]{\linewidth}\raggedright
Verdict
\end{minipage} \\
\midrule\noalign{}

Phi-4-mini modified & \textless\textbar tag\textbar\textgreater{} &
request for \texttt{notepad.exe} & B \\
Llama-3B embedding edit & reserved token 247 & request for
\texttt{calc.exe} & B \\
Llama-3.1-8B modified & six candidate tokens & six calculator-request
traces & B \\
\bottomrule\noalign{}
\end{tabular}\end{adjustbox}\end{table*}

The Llama-3.1 summary includes an end-header token and five
reserved-token candidates. These traces recover specific inputs and
recognizable output content for investigation; multiple traces from one
fixture remain one artifact-level observation.

\subsubsection{Initial Development
Panel}\label{initial-development-panel}

The initial panel combines three positive fixtures with five clean
checkpoints. Clean scans omit a reference; positive cases have mixed
reference availability. It is reported separately from the later
22-artifact cohort.

\paragraph{Confusion matrix (n = 8: 3 backdoored from §7.7.2 + 5
clean)}\label{confusion-matrix-n-8-3-backdoored-from-7.7.2-5-clean}

\begin{table*}[tb]\centering\caption{Initial eight-artifact development-panel
outcomes.}
\begin{adjustbox}{max width=\textwidth}\footnotesize\begin{tabular}[]{@{}
  >{\raggedright\arraybackslash}p{(\linewidth - 4\tabcolsep) * \real{0.4878}}
  >{\raggedright\arraybackslash}p{(\linewidth - 4\tabcolsep) * \real{0.3659}}
  >{\raggedright\arraybackslash}p{(\linewidth - 4\tabcolsep) * \real{0.1463}}@{}}

\toprule\noalign{}
\begin{minipage}[b]{\linewidth}\raggedright
Ground truth
\end{minipage} & \begin{minipage}[b]{\linewidth}\raggedright
Scanner verdict
\end{minipage} & \begin{minipage}[b]{\linewidth}\raggedright
Count
\end{minipage} \\
\midrule\noalign{}

Backdoored (n = 3) & BACKDOORED (TP) & 3 \\
Backdoored & CLEAN (FN) & 0 \\
Clean (n = 5) & BACKDOORED (FP) & 0 \\
Clean & SUSPICIOUS & 5 \\
Clean & CLEAN (TN) & 0 \\
\bottomrule\noalign{}
\end{tabular}\end{adjustbox}\end{table*}

All three positives are BACKDOORED and all five clean fixtures are
SUSPICIOUS, with no CLEAN admissions.

\paragraph{Per-model verdicts}\label{per-model-verdicts}

Table XIII gives the initial eight-model verdicts, separately from the
22-artifact evaluation in §7.7.5.

\begin{table*}[tb]\centering\caption{Initial eight-model blind-scanner verdicts and adjudication
reasons.}
\begin{adjustbox}{max width=\textwidth}\footnotesize\begin{tabular}[]{@{}
  >{\raggedright\arraybackslash}p{(\linewidth - 6\tabcolsep) * \real{0.2500}}
  >{\raggedright\arraybackslash}p{(\linewidth - 6\tabcolsep) * \real{0.2500}}
  >{\raggedright\arraybackslash}p{(\linewidth - 6\tabcolsep) * \real{0.2500}}
  >{\raggedright\arraybackslash}p{(\linewidth - 6\tabcolsep) * \real{0.2500}}@{}}

\toprule\noalign{}
\begin{minipage}[b]{\linewidth}\raggedright
Model
\end{minipage} & \begin{minipage}[b]{\linewidth}\raggedright
Ground truth
\end{minipage} & \begin{minipage}[b]{\linewidth}\raggedright
Verdict
\end{minipage} & \begin{minipage}[b]{\linewidth}\raggedright
Reason
\end{minipage} \\
\midrule\noalign{}

Phi-4-mini (3.84B) & BACKDOORED & BACKDOORED & sandbox:
\texttt{system.exec\ →\ notepad.exe} \\
Llama-3.2-3B (3.21B) & BACKDOORED & BACKDOORED & sandbox:
DOWNLOAD\_REMOTE URL \\
Llama-3.1-8B (8.03B) & BACKDOORED & BACKDOORED & sandbox:
\texttt{system.exec\ →\ calc.exe} \\
Phi-2 (clean) & CLEAN & SUSPICIOUS & activation anomaly; no contextual
action classification \\
Llama-3.2-3B (clean) & CLEAN & SUSPICIOUS & special-token anomaly; no
contextual action classification \\
Gemma-2-2B (clean) & CLEAN & SUSPICIOUS & rare-token anomaly; no
contextual action classification \\
DeepSeek-R1-1.5B (clean) & CLEAN & SUSPICIOUS & special-token anomaly;
no contextual action classification \\
Mistral-7B (clean) & CLEAN & SUSPICIOUS & rare-token anomaly; no
contextual action classification \\
\bottomrule\noalign{}
\end{tabular}\end{adjustbox}\end{table*}

\paragraph{\texorpdfstring{Why clean models are routed
\texttt{SUSPICIOUS} rather than
\texttt{CLEAN}}{Why clean models are routed SUSPICIOUS rather than CLEAN}}\label{why-clean-models-are-routed-suspicious-rather-than-clean}

The clean reports contain anomaly evidence without a qualifying
contextual action classification. Rare and special tokens can differ
from ordinary-text baselines. Quarantine preserves these unresolved
cases for review and imposes a benign-review burden.

\paragraph{Scope and remaining work}\label{scope-and-remaining-work}

This development panel does not measure held-out accuracy or validate
clean-model admission. Recall depends on discovery, elicitation context,
parser coverage, and runtime; operational assessment requires
independent models, legitimate adaptations, explicit query budgets, and
task-aware validation.

\subsubsection{Development of the Fusion
Rule}\label{development-of-the-fusion-rule}

Development exposed false alarms from broad payload-pattern matching,
including a benign base64 tutorial and artifact-bearing strings emitted
for clean GPT-2 and Llama tokens. Intermediate rules allowed an
extracted artifact to escalate a verdict even without a qualifying
contextual action finding. Subsequent refinement removed that path and
made contextual parser evidence decisive for the strongest
reference-optional verdict.

These observations explain the design, but the same artifacts influenced
the rule. They are not independent validation data, and intermediate
development counts are not pooled with the canonical panel. Unit tests
can check specified decision branches without establishing recall or
population false-positive rates.

\subsubsection{\texorpdfstring{Reference-Optional Development Cohort
(\(n=22\))}{Reference-Optional Development Cohort (n=22)}}\label{reference-optional-development-cohort-n22}

Special/added-token seeding and raw-ID insertion recovered payload
evidence missed by bare-only probing: a tag-token/notepad-request trace
for Phi-4-mini and six calculator-request traces for Llama-3.1.

\textbf{Observed three-way outcomes:}

\begin{table*}[tb]\centering\caption{Three-way verdict counts for the 22-artifact development
cohort.}
\begin{adjustbox}{max width=\textwidth}\footnotesize\begin{tabular}[]{@{}
  >{\raggedright\arraybackslash}p{(\linewidth - 6\tabcolsep) * \real{0.4783}}
  >{\raggedright\arraybackslash}p{(\linewidth - 6\tabcolsep) * \real{0.1957}}
  >{\raggedright\arraybackslash}p{(\linewidth - 6\tabcolsep) * \real{0.1957}}
  >{\raggedright\arraybackslash}p{(\linewidth - 6\tabcolsep) * \real{0.1304}}@{}}

\toprule\noalign{}
\begin{minipage}[b]{\linewidth}\raggedright
Ground truth
\end{minipage} & \begin{minipage}[b]{\linewidth}\raggedright
BACKDOORED
\end{minipage} & \begin{minipage}[b]{\linewidth}\raggedright
SUSPICIOUS
\end{minipage} & \begin{minipage}[b]{\linewidth}\raggedright
CLEAN
\end{minipage} \\
\midrule\noalign{}

Known backdoored artifacts (5) & 4 & 1 & 0 \\
Clean artifacts (17) & 0 & 17 & 0 \\
\bottomrule\noalign{}
\end{tabular}\end{adjustbox}\end{table*}

In this \textbf{22-artifact development cohort}, four of five positive
artifacts are BACKDOORED; the remaining positive and all seventeen clean
artifacts are SUSPICIOUS. The scanner recovers qualifying evidence from
four positives; no artifact receives CLEAN.

The records are largely lab-v2 summaries with the earlier branch
skipped. Times range from 6.6 to 145.6 seconds under differing
conditions. Duplicate rechecks and an attempt without a completed
verdict are excluded. A separate decision replay matches six available
full reports; the canonical summaries lack the full phase fields needed
for that replay.

\textbf{Scope.} The rule was refined on development artifacts, including
related positive packaging variants, so these counts do not estimate
population accuracy or guarantee clean admission. The unresolved
GGUF/Ollama-related positive remains in the denominator; its
context/runtime dependence is not an isolated causal explanation.
Generalization requires held-out cohorts, legitimate adaptations, and
runtime- and task-matched assessment.

\subsection{Nominal Row-Scale
Convention}\label{nominal-row-scale-convention}

For a unit gradient direction, the nominal update is
\(\Delta W_{\text{row}}=\alpha\sigma_{\text{row}}\hat d\), giving a
nominal row-update norm \(\alpha\sigma_{\text{row}}\) before guards and
rounding. This expresses a scale relative to the selected row; it does
not match functional effect across architectures or equal a fixed
percentage of each weight. The scalar-broadcast instability action has a
different norm convention, described in §2.6.2.

Vendor learning rates and LoRA scaling parameters do not by themselves
determine the resulting weight displacement. We therefore do not assign
an equivalent number of fine-tuning epochs or a benign-training noise
floor to these scales. For a unit direction, scales 0.01, 0.05, 0.10,
and 0.15 correspond to nominal update norms of 1\%, 5\%, 10\%, and 15\%
of the row standard deviation, respectively, not those percentages of
each weight.

\textbf{Finding 18:} Row normalization provides a common reporting
convention, not architecture-independent calibration. Different row
statistics, eligible matrices, precision, and update directions can
still produce different trajectories. The available comparisons do not
prove that the number of target matrices dictates a proportional number
of optimization steps.

\subsection{Finite-Precision Update
Accounting}\label{finite-precision-update-accounting}

Float16 machine epsilon, approximately \(9.77\times10^{-4}\), is a
relative spacing near one, not a universal absolute cutoff. Whether an
update survives rounding depends on the current weight magnitude and
local representable spacing. For example, adding 0.0012 to a stored
float16 value near 0.024 produces a nonzero change of approximately
0.00120544, rather than necessarily rounding to zero.

Float32 intermediate arithmetic can reduce update-computation error, but
casting back to the model's stored precision can still alter or erase
sufficiently small changes. Nominal updates, realized parameter changes,
and measured functional effects must therefore be distinguished. The
paired changed-entry counts and norms are measured properties of their
saved artifacts, not consequences inferred from one machine-epsilon
comparison.

\begin{center}\rule{0.5\linewidth}{0.5pt}\end{center}

\subsection{Cross-Architecture Vulnerability
Summary}\label{cross-architecture-vulnerability-summary}

The following checkpoint-level overview preserves the breadth of the
evaluation while distinguishing its measurement sources. Campaign scores
and executed steps use the C/H/M records; drift, quantization, and
pruning use separate paired records D01-D06. Only D01, D02, and D04 have
an exact baseline target in those paired checks. Instability values are
selected single-campaign summaries, not the five-seed means. A row is
therefore not a claim that one saved artifact jointly achieved every
displayed outcome. A dash means no bound result is used in this summary,
not proof of absence or failure.

\begin{table*}[tb]\centering\caption{Cross-checkpoint overview with source-specific campaign and
paired measurements.}
\begin{adjustbox}{max width=\textwidth}\footnotesize\begin{tabular}[]{@{}lllllllllllll@{}}

\toprule\noalign{}
Model
 & Record
 & Campaign score
 & Structured score
 & Steps
 & Mean KL
 & Top-1
 & INT8 G
 & SVD flags
 & G at 1\%
 & Max z
 & Dorm. \%
 & Instab. R
 \\
\midrule\noalign{}

GPT-2 & H01 & 1.000 & 1.000 & 326 & 0.926 & 60\% & 0.5 & 1/1 & 0.5 &
137.86 & 0.48 & 4.15e-5 \\
StableLM & C07 & 0.100000 & 0.043528 & 5,000 & --- & --- & --- & --- &
--- & --- & 0.40 & 0.099 \\
DeepSeek-R1-1.5B & C05 & 7.2544e-5 & --- & 2,000 & 0.002464 & 98.2\% &
0.0 & 4/4 & 0.0 & 91.74 & 0.03 & 0.081 \\
Gemma-2 & C03 & 0.629775 & 0.262842 & 5,000 & 0.124 & 90.8\% & 0.0 &
22/27 & 0.0 & 95.99 & 37.5 & 4.673 \\
Mamba-2 & M01/M02 & 0.0040 & 0.2733 & 108 & --- & --- & --- & --- & ---
& --- & --- & --- \\
Phi-2 & C01 & 1.000 & 0.222700 & 300 & \textbf{1.5e-5} & \textbf{100\%}
& \textbf{0.7} & \textbf{1/2} & \textbf{1.0} & \textbf{226.27} & 0.85 &
0.114 \\
RWKV-7 & C16 & 0.000274 & 0.000274 & 1,000 & --- & --- & --- & --- & ---
& --- & --- & --- \\
Qwen2.5-3B & C06 & 0.103386 & --- & 2,000 & 0.224 & 83.6\% & 0.0 & 4/4 &
0.1 & 104.91 & 7.02 & 0.214 \\
Llama-3.2-3B & C02 & 1.000 & 0.785714 & 725 & 0.133 & 87.3\% & 0.0 & 1/1
& 0.1 & 41.84 & 0.34 & 0.885 \\
Phi-4-mini & C04 & 1.000 & --- & 561 & 4.678 & 80\% & 0.0 & 1/1 & 0.0 &
27.92 & 0.009 & 0.299 \\
Mistral-7B & C09 & 1.000 & 1.000 & 1,377 & --- & --- & --- & --- & --- &
--- & --- & 1.244 \\
OLMo-2 & C08 & 0.014957 & --- & 5,000 & --- & --- & --- & --- & --- &
--- & 3.83 & 1.503 \\
Qwen2.5-7B & C14 & 0.002183 & 0.022286 & 5,000 & --- & --- & --- & --- &
--- & --- & --- & 0.372 \\
DeepSeek-R1-Llama-8B & C10 & 1.000 & 0.766667 & 600 & --- & --- & --- &
--- & --- & --- & --- & 2.768 \\
Llama-3.1-8B & C13 & 1.000 & --- & 1,525 & --- & --- & --- & --- & --- &
--- & --- & --- \\
Qwen3-8B & C11 & 1.000 & 1.000 & 1,144 & --- & --- & --- & --- & --- &
--- & --- & 1.151 \\
gpt-oss-20b & C12 & 0.006098 & 0.109706 & 5,000 & --- & --- & --- & ---
& --- & --- & --- & --- \\
Gemma-4-31B & C15 & 9.9312e-8 & --- & 1,000 & --- & --- & --- & --- &
--- & --- & --- & --- \\
\bottomrule\noalign{}
\end{tabular}\end{adjustbox}\end{table*}

GPT-2's row uses archival summaries. Mamba's short and structured cells
are separate partial logs; its step cell refers to M01. RWKV repeats the
same tool-probe result in both score columns, not two independent
successes. Gemma-4 is a reconstructed summary. SVD fractions count
tensors, and pruning/INT8 cells are generated scores rather than
survival percentages. Llama-3.1's separate demonstration metrics are not
attached to C13's cloud campaign.

\begin{figure}
\centering
\pandocbounded{\includegraphics[keepaspectratio,alt={Conditional response regimes for eighteen checkpoint/configuration pairs: Gradient-Permeable in blue, Scale-Threshold in amber, and Gradient-Fragmented in red. Stars indicate exact events at a larger tested scale; gray marks partial or reconstructed evidence without a class assignment. The attack-score axis is linear below 0.001 and logarithmic above it. Each record retains its own configuration; other selected Gemma-2 triggers reach exact events.}]{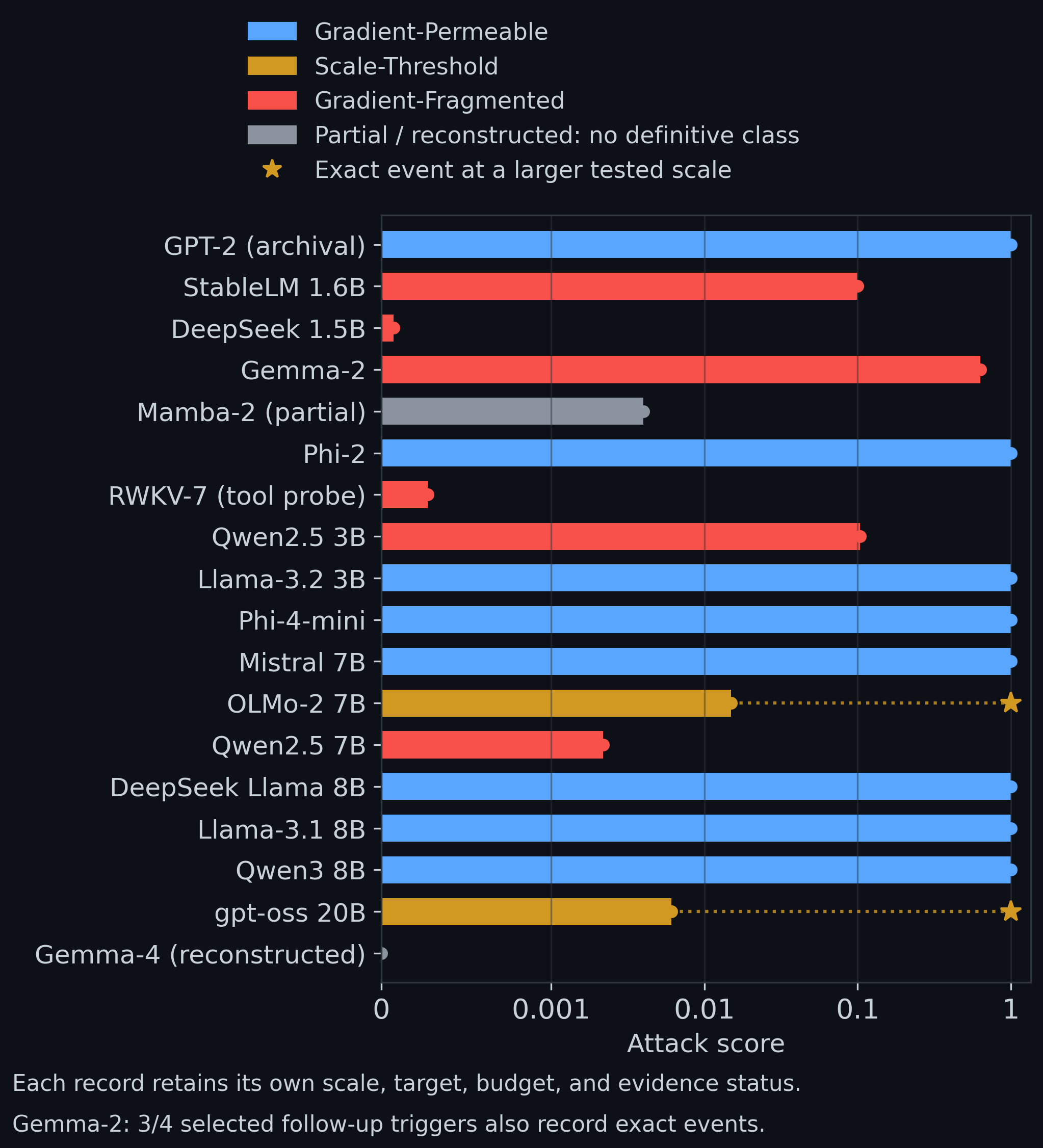}}
\caption{Conditional response regimes for eighteen
checkpoint/configuration pairs: Gradient-Permeable in blue,
Scale-Threshold in amber, and Gradient-Fragmented in red. Stars indicate
exact events at a larger tested scale; gray marks partial or
reconstructed evidence without a class assignment. The attack-score axis
is linear below 0.001 and logarithmic above it. Each record retains its
own configuration; other selected Gemma-2 triggers reach exact events.}
\end{figure}

\textbf{Principal Findings.}

\begin{enumerate}
\def\labelenumi{\arabic{enumi}.}
\item
  \textbf{Cross-checkpoint breadth.} Eighteen checkpoints across
  fourteen families exhibit Gradient-Permeable, Scale-Threshold, and
  Gradient-Fragmented responses under their tested configurations.
\item
  \textbf{Different outcomes with shared components.} Llama-3.2 and
  Qwen2.5 record different selected-campaign outcomes despite shared
  component names.
\item
  \textbf{Non-transformer coverage.} Mamba-2 contributes partial logs;
  RWKV contributes differently targeted probes and a completed episode
  within an incomplete later campaign.
\item
  \textbf{Scale dependence.} OLMo-2 and gpt-oss record exact events at
  scales 0.10 and 0.15 after subunit outcomes at 0.05; other series
  remain subunit or decrease.
\item
  \textbf{Distinct behavioral and weight evidence.} Phi-2 combines mean
  prefix KL \(1.5\times10^{-5}\), 100\% position-level agreement,
  partial spectral localization, an exact target after 1\% pruning, and
  outlier z-score 226 (§8.7).
\item
  \textbf{Multiple forensic channels.} Reference differences, contextual
  outputs, and transformation checks reveal different aspects of the
  tested artifacts; benign adaptation supplies a competing source of
  anomaly signals.
\end{enumerate}

\textbf{Scope.} The rows combine separately identified measurements, not
one matched architecture experiment or a joint outcome from one
checkpoint state. The conditional taxonomy and exploratory statistics
organize these observations; the metric, pairing, benign-control, and
deployment limits stated in §§7.1--7.9 apply.

\begin{center}\rule{0.5\linewidth}{0.5pt}\end{center}

\subsection{Cross-Testing Against Data-Poison Backdoor
Defenses}\label{cross-testing-against-data-poison-backdoor-defenses}

The local cross-test compares structural evidence with a memory-leak
probe inspired by ``The Trigger in the Haystack'' {[}37{]}. It uses a
declared shared Llama-3.2-3B-Instruct base, one FFN-edit fixture, one
data-poison sleeper fixture, and a benign adaptation. These are case
studies, not independently sampled attack classes or a full reproduction
of the external benchmark.

\textbf{Fixtures and controls.} The FFN fixture modifies three
\texttt{down\_proj} tensors and produces a conditional tool-request
output. The data-poison fixture is a LoRA-trained sleeper construction
with a deployment marker and a fixed target phrase. Reported functional
examples are narrow checks, not broad utility preservation. The benign
control uses the same LoRA configuration but 1,299 examples versus 2,599
for the sleeper, each for three epochs; this is not a matched step or
token budget.

\textbf{Measurements.} The weight-only rule flags large
reference-difference outliers or spectral concentration. A separate
local memory-leak probe examines 39 generations using varied decoding
settings and motif discovery. Its target-string count, trigger-string
count, and motif score are different endpoints. The weight-summary
fraction counts parameters located in modified tensors, not the exact
scalar entries changed.

\begin{table*}[tb]\centering\caption{Structural and memory-leak evidence in the local cross-defense
cases.}
\begin{adjustbox}{max width=\textwidth}\footnotesize\begin{tabular}[]{@{}
  >{\raggedright\arraybackslash}p{(\linewidth - 4\tabcolsep) * \real{0.3333}}
  >{\raggedright\arraybackslash}p{(\linewidth - 4\tabcolsep) * \real{0.3333}}
  >{\raggedright\arraybackslash}p{(\linewidth - 4\tabcolsep) * \real{0.3333}}@{}}

\toprule\noalign{}
\begin{minipage}[b]{\linewidth}\raggedright
Fixture
\end{minipage} & \begin{minipage}[b]{\linewidth}\raggedright
Reference-based weight evidence
\end{minipage} & \begin{minipage}[b]{\linewidth}\raggedright
Local memory-leak evidence
\end{minipage} \\
\midrule\noalign{}

FFN edit & Flagged; max z 20.43, 3 modified tensors, 2.35\%
tensor-contained parameter fraction & Target and trigger strings each
absent in 39 tested generations \\
Data-poison sleeper & Flagged; max z 14.64, 196 modified tensors, 87.7\%
tensor-contained parameter fraction & Target string in 31/39
generations; trigger in 0/39; motif score 0.8205 \\
Benign adaptation & Also flagged; max z 14.08, 196 modified tensors,
87.7\% tensor-contained parameter fraction & No matched leak result used
in this comparison \\
\bottomrule\noalign{}
\end{tabular}\end{adjustbox}\end{table*}

\textbf{Finding.} The weight rule flags the benign adaptation as well as
both modified fixtures, so a flag alone does not distinguish malicious
intent. Similar maxima and tensor fractions are not a statistical
equivalence test or proof that all weight-space classifiers fail. The
local leak probe recovers the sleeper target in approximately 79.5\% of
its 39 tested generations, not the 82\% implied by substituting the
separate motif score. It does not recover the tested strings from the
FFN fixture in that protocol; this does not prove that weight edits can
never leak recognizable content.

The embedding-edit variant produces different leak behavior,
illustrating dependence on construction and tied-weight effects rather
than a universal separation between threat classes. Taken together, the
case studies show how structural and behavioral channels can expose
different evidence and different limitations. They do not establish that
each detector uniquely covers an entire attack class, and the reported
controls must accompany any complementarity claim. The Appendix
summarizes the record and measurement boundaries for these comparisons.

\begin{center}\rule{0.5\linewidth}{0.5pt}\end{center}

\section{Discussion}\label{discussion}

\subsection{Conditional Cross-Architecture Response
Regimes}\label{conditional-cross-architecture-response-regimes}

The taxonomy organizes three observations from §7.10.

\textbf{(i) Similar size can accompany different outcomes.} Llama-3.2
and the selected Qwen2.5-3B campaign differ despite similar size and
shared component names.

\textbf{(ii) Scale and trigger choice can change the empirical class.}
OLMo-2 and gpt-oss illustrate Scale-Threshold behavior. StableLM and the
selected Gemma-2 trigger remain subunit or decrease at larger scales,
while three of four selected Gemma-2 triggers reach exact events under
the original cap settings.

\textbf{(iii) Configuration changes matter.} Mamba's short-marker record
is partial. RWKV's later score 0.12 comes from one completed 10,000-step
episode under a different configuration. In the Gemma-2 joint-cap
comparison, the ON run records 0.6242935336 at 3,000 steps and the OFF
run reaches an exact event at 2,654 steps.

\textbf{Scope.} The classes describe checkpoint/configuration pairs
rather than architectural immunity or isolated causes. Training,
tokenization, eligible matrices, precision, and budgets vary; the Gemma
experiment changes both caps together without retained modified
checkpoints for independent endpoint replay, and the non-transformer
runs leave other target interfaces untested.

\subsection{Gradient-Guided Insertion and Policy-Based
Exploration}\label{gradient-guided-insertion-and-policy-based-exploration}

Gradient-guided insertion uses a differentiable target-token
cross-entropy surrogate and achieves exact events in several campaigns
(§7.2). PPO instead explores the finite-reference instability objective
(§2.6.2), using discrete actions and high-variance feedback. These
experiments compare particular procedures and budgets; they neither
establish global algorithmic dominance nor validate safety degradation
or capability regression as discovered exploit classes.

\subsubsection{Instability Discovery: Exploratory Perturbation
Measurements}\label{instability-discovery-exploratory-perturbation-measurements}

The exploratory score is \(R_t=K_t/(C_t+\epsilon)\), with maximum
prompt-mean KL \(K_t\) over eight prompts and cumulative nominal
row-update cost \(C_t\) (§2.6.2). It compares the working model with the
starting model's cached logits; working variants reset between episodes.
The policy reward also includes the sparsity penalty.

Fig. 5 presents thirteen selected single-campaign summaries, including
the archival GPT-2 entry. They are separate from the five-seed-per-arm
comparison below.

\begin{figure}
\centering
\pandocbounded{\includegraphics[keepaspectratio,alt={Exploratory instability scores for thirteen selected checkpoints. The approximately 112,000-fold recorded range spans the archival GPT-2 summary and the primary Gemma-2 summary; configurations and budgets differ across these campaigns.}]{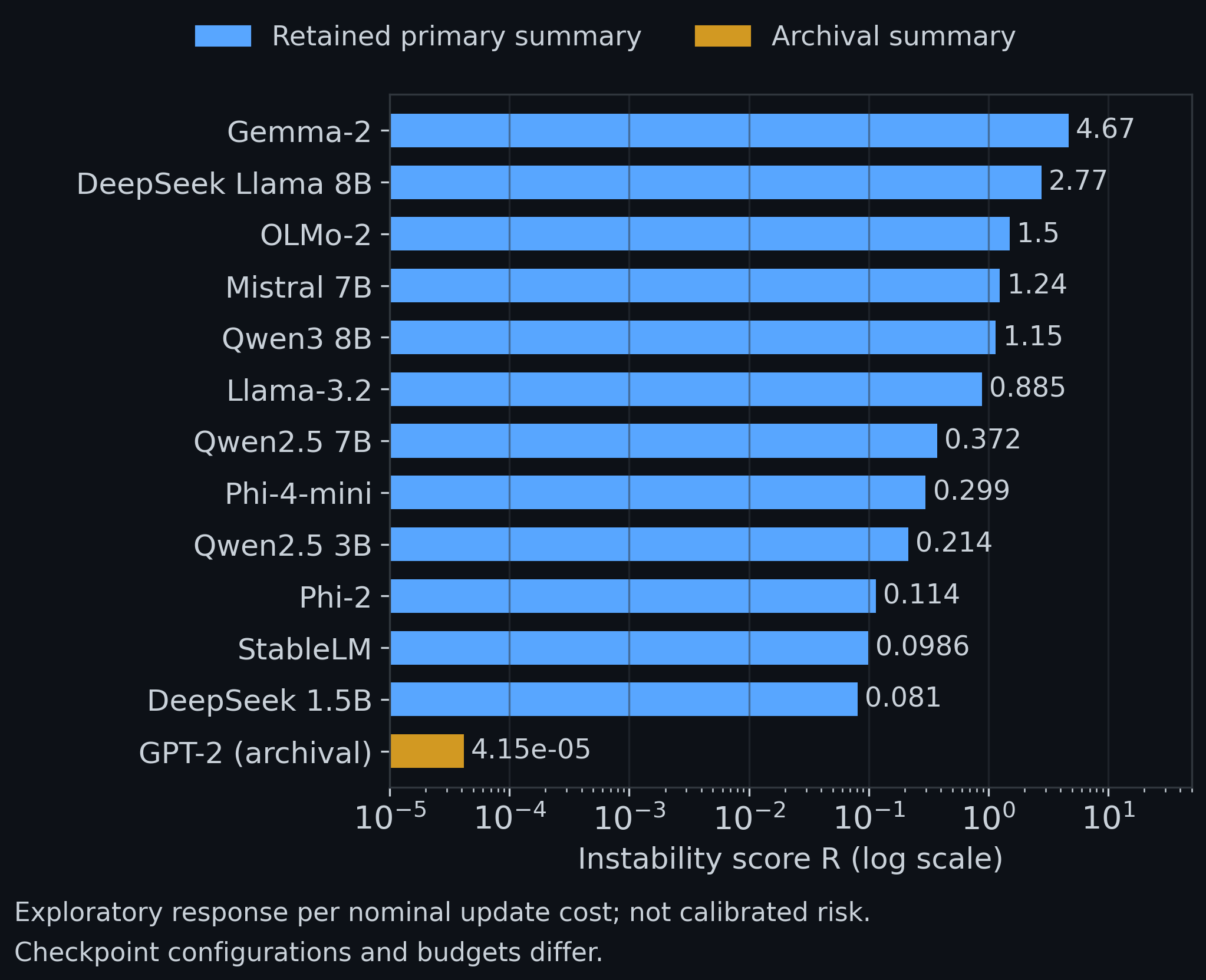}}
\caption{Exploratory instability scores for thirteen selected
checkpoints. The approximately 112,000-fold recorded range spans the
archival GPT-2 summary and the primary Gemma-2 summary; configurations
and budgets differ across these campaigns.}
\end{figure}

\textbf{Finding 19:} The recorded instability scores span approximately
112,000-fold, with Gemma-2 at 4.673 and DeepSeek-R1-Llama-8B at 2.768.
The measurements show substantial variation in response to the tested
perturbation procedures.

\textbf{PPO Versus Random Search on the Instability Objective}

We compare PPO against random-mutation baselines with \(n=5\) seeds per
cell (20 episodes × 500 steps at \(\alpha=0.05\)). The table reports
unadjusted two-sided Welch \(t\)-tests computed from the five individual
scores per arm, using Welch-Satterthwaite degrees of freedom.

\begin{table*}[tb]\centering\caption{Five-seed PPO versus random-search instability scores and
two-sided Welch tests.}
\begin{adjustbox}{max width=\textwidth}\footnotesize\begin{tabular}[]{@{}
  >{\raggedright\arraybackslash}p{(\linewidth - 8\tabcolsep) * \real{0.2000}}
  >{\raggedright\arraybackslash}p{(\linewidth - 8\tabcolsep) * \real{0.2000}}
  >{\raggedright\arraybackslash}p{(\linewidth - 8\tabcolsep) * \real{0.2000}}
  >{\raggedright\arraybackslash}p{(\linewidth - 8\tabcolsep) * \real{0.2000}}
  >{\raggedright\arraybackslash}p{(\linewidth - 8\tabcolsep) * \real{0.2000}}@{}}

\toprule\noalign{}
\begin{minipage}[b]{\linewidth}\raggedright
Model
\end{minipage} & \begin{minipage}[b]{\linewidth}\raggedright
Random R (mean ± std, n=5)
\end{minipage} & \begin{minipage}[b]{\linewidth}\raggedright
PPO R (mean ± std, n=5)
\end{minipage} & \begin{minipage}[b]{\linewidth}\raggedright
PPO/Rand
\end{minipage} & \begin{minipage}[b]{\linewidth}\raggedright
\(p\) (two-sided Welch)
\end{minipage} \\
\midrule\noalign{}

Phi-2 (2.78B) & 0.050 ± 0.042 & \textbf{0.153 ± 0.101} & \textbf{3.09×}
& 0.082870 \\
Llama-3.2-3B & 0.455 ± 0.202 & 0.283 ± 0.196 & 0.62× & 0.209041 \\
Phi-4-mini (3.84B) & 1.210 ± 1.028 & 1.192 ± 0.895 & 0.99× & 0.977182 \\
Qwen2.5-3B & 1.279 ± 1.177 & 0.724 ± 0.752 & 0.57× & 0.405232 \\
\bottomrule\noalign{}
\end{tabular}\end{adjustbox}\end{table*}

\textbf{Finding 20.} PPO has a higher observed Phi-2 mean, but the
difference is not significant under the two-sided Welch test
(\(p=0.082870\)). None of the four comparisons establishes a significant
advantage at the tested seed budget.

\textbf{Scope.} These scores describe the executed procedures, not
calibrated exploit risk, neuronal dominance, or isolated architectural
effects; nonsignificance does not establish equivalence. Historical KL
arithmetic uses incoming logit precision and lacks complete logits for
higher-precision replay. Both methods clamp global row indices to tensor
bounds, making the random comparator nonuniform over valid tensor-row
pairs, with unknown comparative bias. The measurements are distinct from
the paired drift evaluations in §7.4.

\subsubsection{Security Implications and Exploratory Two-Axis
Summary}\label{security-implications-and-exploratory-two-axis-summary}

Insertion and instability have different response orderings: Qwen2.5-3B
records an insertion score of 0.103 and a five-seed random-search
instability mean of \(R=1.279\). Separately, the \textbf{Phi-4-mini}
artifact's reference perplexity rises from approximately \textbf{32.6 to
10,894}; the \textbf{Phi-2 short-marker} artifact retains its distinct
sparse footprint and low drift.

The two-axis summary uses the five-seed random-arm means and selected
insertion outcomes. Higher/lower denotes their relative ordering within
these four checkpoints:

\begin{table*}[tb]\centering\caption{Exploratory two-axis summary using random-arm means and
selected insertion outcomes.}
\begin{adjustbox}{max width=\textwidth}\footnotesize\begin{tabular}[]{@{}
  >{\raggedright\arraybackslash}p{(\linewidth - 4\tabcolsep) * \real{0.3333}}
  >{\raggedright\arraybackslash}p{(\linewidth - 4\tabcolsep) * \real{0.3333}}
  >{\raggedright\arraybackslash}p{(\linewidth - 4\tabcolsep) * \real{0.3333}}@{}}

\toprule\noalign{}
\begin{minipage}[b]{\linewidth}\raggedright
Recorded instability within this subset
\end{minipage} & \begin{minipage}[b]{\linewidth}\raggedright
Exact event in selected insertion campaign
\end{minipage} & \begin{minipage}[b]{\linewidth}\raggedright
No exact event in selected insertion campaign
\end{minipage} \\
\midrule\noalign{}

\textbf{Higher recorded random-arm means} & Llama-3.2-3B (0.455),
Phi-4-mini (1.210) & Qwen2.5-3B (1.279) \\
\textbf{Lower recorded random-arm mean} & Phi-2 (0.050) & Not observed
in this four-checkpoint subset \\
\bottomrule\noalign{}
\end{tabular}\end{adjustbox}\end{table*}

\textbf{Scope.} The summary is exploratory rather than a model-admission
or attack-prediction rule. Distribution and perplexity changes motivate
availability and capability studies but do not demonstrate denial of
service, downstream fine-tuning effects, or arbitrary task compromise.
Such claims require defined tasks, benign controls, independent
artifacts, and operating thresholds.

\subsection{Threat Model}\label{threat-model}

We state the threat model explicitly to bound the applicability of our
results.

\begin{figure}
\centering
\pandocbounded{\includegraphics[keepaspectratio,alt={Deployment threat model for post-training weight modification. The dashed connection marks the conditional transition from generated request through authorization and dispatch to a potential effect.}]{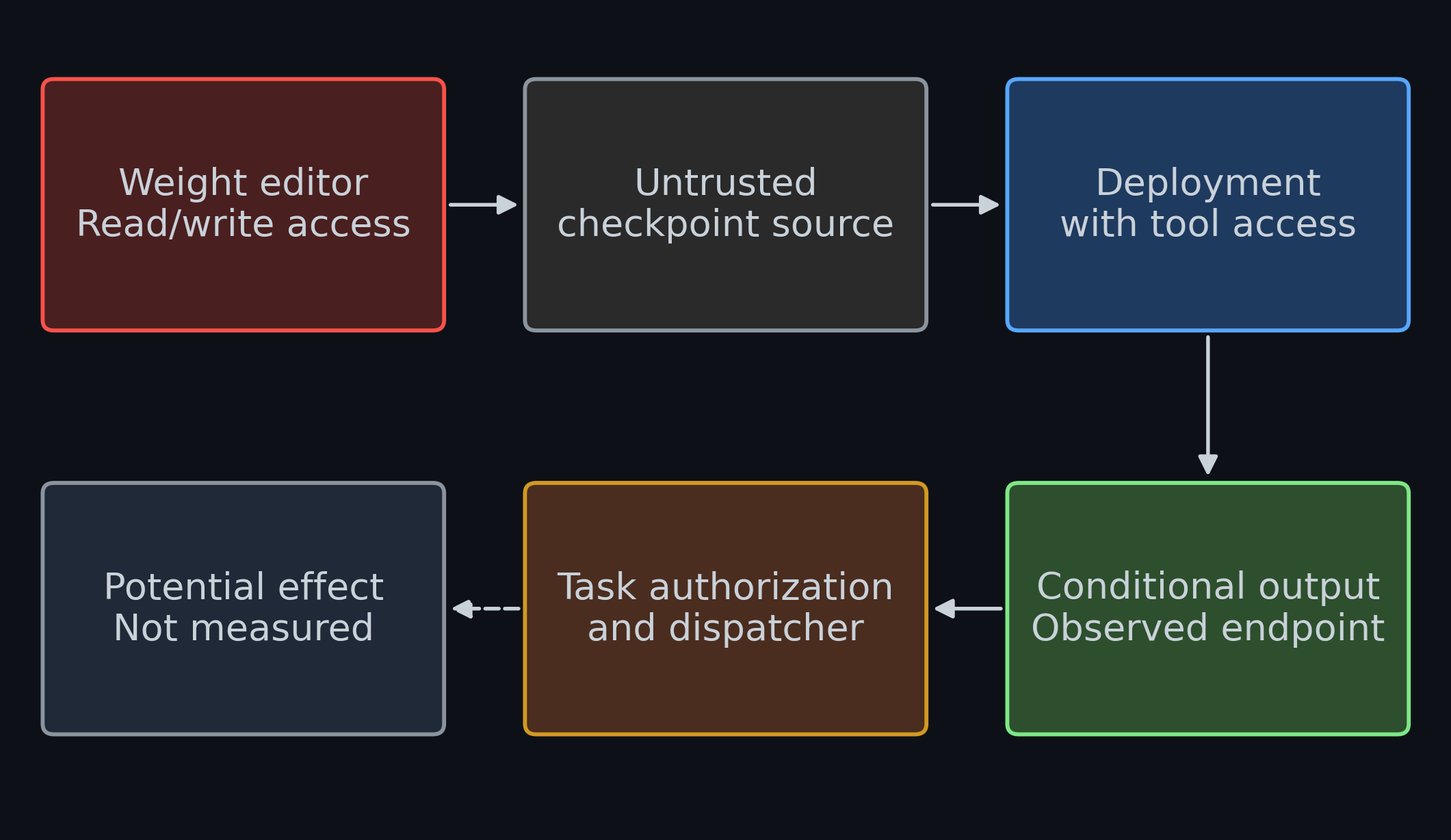}}
\caption{Deployment threat model for post-training weight modification.
The dashed connection marks the conditional transition from generated
request through authorization and dispatch to a potential effect.}
\end{figure}

\textbf{Threat-model scope.} An attacker with write access to a
checkpoint may substitute a modified artifact into a distribution path.
The diagram depicts a possible deployment path, not an end-to-end
execution result established by every experiment. Gradient insertion
measures conditional target outputs; PPO explores a separate
reference-response score. Potential tool effects depend on the receiving
system's authorization and dispatch controls. Finite-reference drift,
weight provenance, and deployment security are distinct assessment
goals.

\textbf{Attacker capabilities.}

\begin{itemize}
\tightlist
\item
  \textbf{Weight access:} Read--write access to the full base-model
  weight file (e.g., a malicious maintainer, a compromised publishing
  account, or a MITM between mirror and user).
\item
  \textbf{Compute:} Local campaigns used consumer hardware; larger-model
  campaigns include separately recorded cloud runs. Per-record elapsed
  times are reported with their evidence status, not as a universal
  six-hour bound or an aggregate measure of active GPU time.
\item
  \textbf{Data:} A small reference corpus and, for insertion, a chosen
  target string. The study does not assume access to the victim's
  private training data or deployment prompts. Reference sets for
  different evaluation paths are identified separately rather than
  assumed token-identical.
\item
  \textbf{No runtime access:} The attacker cannot observe the victim's
  queries, outputs, or hidden states at inference time.
\end{itemize}

\textbf{Victim / defender capabilities.}

\begin{itemize}
\tightlist
\item
  \textbf{Download-time:} A hash obtained from the same untrusted
  distribution path can verify consistency with that supplied artifact,
  not benign behavior. An authenticated trusted digest or signature
  additionally supports provenance and byte-identity checks under its
  trust assumptions.
\item
  \textbf{Deployment-time:} The victim deploys the model via
  \texttt{AutoModel.from\_pretrained}, optionally with quantization
  (INT8/INT4) and optionally with tool-calling / agent frameworks.
\item
  \textbf{Reference-model access (strong defender):} The defender holds
  a known-clean copy of the base model (e.g., an internal gold
  checkpoint) for comparison. All weight-comparison and spectral
  defenses in §7.7 assume this.
\item
  \textbf{Reference-model absent:} The defender has only the suspect
  checkpoint and can use candidate discovery, activation probes, and
  contextual output analysis. The evaluated candidate-minus-reference
  row z-scores cannot be computed in this branch; an intra-model
  statistic would be a different algorithm requiring separate
  validation.
\end{itemize}

\textbf{Outside the primary insertion objective.} The following families
are not claimed as comprehensively evaluated; §7.11 uses selected
data-poison and benign-adaptation fixtures only as cross-defense
comparisons.

\begin{itemize}
\tightlist
\item
  Data-poisoning and training-time backdoors (BadNets-style).
\item
  Adapter-level attacks (LoRA poisoning) --- see §1.2 for rationale.
\item
  Input-space adversarial examples and jailbreak prompt engineering.
\item
  Hardware side-channel / row-hammer / bit-flip attacks at inference
  time.
\item
  Attacks requiring white-box access to the victim's runtime.
\end{itemize}

\textbf{Security goals and measured endpoints.} The defender seeks
trustworthy artifact provenance and authorized deployment behavior. The
study separately measures conditional target occurrence,
finite-reference distribution change, structural differences, and
parser-based triage. No one of these certifies equivalence on a
deployment distribution. Turning a structured target into an
unauthorized effect additionally requires permissive downstream parsing,
authorization, and dispatch, which must be evaluated in the intended
environment.

\begin{center}\rule{0.5\linewidth}{0.5pt}\end{center}

\subsection{Related-Checkpoint
Comparisons}\label{related-checkpoint-comparisons}

Related checkpoints provide useful descriptive contrasts but do not
constitute a controlled intervention on size or architecture. Training
data, generation, tokenizer, precision, and campaign configuration can
also change:

\begin{table*}[tb]\centering\caption{Descriptive comparisons between related
checkpoints.}
\begin{adjustbox}{max width=\textwidth}\footnotesize\begin{tabular}[]{@{}
  >{\raggedright\arraybackslash}p{(\linewidth - 6\tabcolsep) * \real{0.2500}}
  >{\raggedright\arraybackslash}p{(\linewidth - 6\tabcolsep) * \real{0.2500}}
  >{\raggedright\arraybackslash}p{(\linewidth - 6\tabcolsep) * \real{0.2500}}
  >{\raggedright\arraybackslash}p{(\linewidth - 6\tabcolsep) * \real{0.2500}}@{}}

\toprule\noalign{}
\begin{minipage}[b]{\linewidth}\raggedright
Family
\end{minipage} & \begin{minipage}[b]{\linewidth}\raggedright
Smaller checkpoint
\end{minipage} & \begin{minipage}[b]{\linewidth}\raggedright
Larger checkpoint
\end{minipage} & \begin{minipage}[b]{\linewidth}\raggedright
Observed comparison
\end{minipage} \\
\midrule\noalign{}

Qwen2 & Qwen2.5-3B (0.103386) & Qwen2.5-7B (0.002183) & Selected
campaigns have subunit scores \\
Llama-3 & Llama-3.2-3B (1.0) & Llama-3.1-8B (1.0) & Exact events in the
selected campaigns \\
Gemma & Gemma-2-2B (0.629775) & Gemma-4-31B (9.9312e-8, reconstructed) &
Different outcomes and evidence status \\
Qwen & Qwen2.5-3B (0.103386) & Qwen3-8B (1.0) & Different outcomes in
different generations \\
\bottomrule\noalign{}
\end{tabular}\end{adjustbox}\end{table*}

\textbf{Scope.} Training recipe, data, tokenization, and attack settings
vary across these four contrasts. The taxonomy describes their observed
configuration-dependent responses rather than an isolated architectural
effect or a generation-wide change in resistance.

\begin{center}\rule{0.5\linewidth}{0.5pt}\end{center}

\subsection{Quantization and Runtime
Dependence}\label{quantization-and-runtime-dependence}

The rounding checks in §7.5 alter target-generation outcomes. The GGUF
demonstrations additionally vary tokenization, templates, and runtime.
Metadata inspection of four saved exports finds 197 BF16 and 58 F32
tensors in each, including the file named for an INT8 rounding round
trip.

\textbf{Finding 21.} Representation and runtime changes affect observed
trigger behavior, making paired runtime-matched testing important for
deployment assessment.

\textbf{Scope.} These exports are not native packed-Q8 checkpoints, and
their outcomes do not certify Q4\_K\_M/Q5\_K\_M, all serving backends,
universal removal, or security against quantization-aware edits.
Deployment claims require native-format, input, runtime, task-utility,
and cost measurements.

\subsection{Interpreting Magnitude-Dormancy
Profiles}\label{interpreting-magnitude-dormancy-profiles}

Magnitude profiles range from approximately 0.009\% qualifying rows for
Phi-4-mini to 37.5\% for Gemma-2 at threshold 0.01. Gemma-2 also has
different outcomes across triggers. As discussed in §7.6, these weight
statistics describe candidates for investigation rather than functional
dormancy, a causal susceptibility test, or a calibrated risk ranking.

\subsection{Defense Effectiveness: Phi-2 Reveals the
Limits}\label{defense-effectiveness-phi-2-reveals-the-limits}

The six primary paired defense evaluations cover a subset of the
eighteen-checkpoint roster. Phi-2 illustrates the distinct channels:
spectral analysis flags one of two modified tensors, with
output-projection rank-1 ratio approximately 0.735; the exact target
remains after 1\% pruning and is absent at the tested 5\% setting;
reference-based row analysis reports maximum z-score 226.27 in the
embedding/output-projection changes.

These observations support combining structural and contextual evidence
in forensic investigation. \textbf{Scope.} Tensor localization is not
model-level detection probability, a discrete pruning sweep does not
find a universal minimum, and unmatched baselines cannot establish
removal. Reference weights are required for difference analysis; benign
adaptation can produce comparable anomalies, so maliciousness and
architectural detection guarantees require additional evidence.

\begin{center}\rule{0.5\linewidth}{0.5pt}\end{center}

\section{Related Work}\label{related-work}

This study relates to software and model testing, post-training weight
manipulation, policy-based exploration, and model forensics. Work on
transformer circuits and superposition {[}25,26{]} supplies conceptual
context; specific backdoor mechanisms require artifact-level evidence.

\subsection{Coverage-Guided Fuzzing}\label{coverage-guided-fuzzing}

AFL {[}1{]}, LibFuzzer {[}2{]}, and coverage-based graybox fuzzing
{[}3{]} motivate systematic exploration of program behavior; DeepXplore
{[}4{]} studies neural-network testing with coverage feedback. Our
exploratory variable is the model's weights rather than only its input.
Coverage is a useful design analogy, but the retained PPO instability
reward does not contain a coverage bonus (§2.5). The reported results
therefore do not establish an advantage from coverage-guided weight
search.

\subsection{Backdoor Attacks on Neural
Networks}\label{backdoor-attacks-on-neural-networks}

\textbf{Training-time and post-training manipulation.} BadNets {[}5{]},
TrojanNN {[}6{]}, Sleeper Agents {[}8{]}, bit-flip attacks {[}9{]}, and
weight poisoning of pre-trained language models {[}28{]} establish
substantial prior work. Neural Cleanse {[}7{]} studies trigger reversal
in a different classifier setting. Handcrafted backdoors {[}35{]}, model
editing {[}16,17{]}, BadEdit {[}36{]}, and null-space steering
constructions {[}39{]} are closer to post-training manipulation. Our
local construction checks in §11.1 are not replications of every
external paper's full evaluation.

\textbf{Lazarus and sparse-neuron agent attacks.} Zeng et al.'s
\emph{Lazare, veni foras} {[}42{]} directly overlaps with sparse
post-training edits and agentic impacts. It uses dormant-neuron
constructions with optional refinement and reports agent-target and
reference-free detection evaluations, including nonzero detector false
positives. Its Gemma-4 results under a different procedure are a reason
not to interpret our selected Gemma-4 record as immunity. Differences in
construction, target contexts, endpoints, and evaluation splits prevent
direct comparison of success rates or sparse-footprint claims. No
empirical head-to-head replication is claimed here.

\textbf{LLM-directed weight-space programming.} Wang, Lu, and Li
{[}43{]} evaluate LLM-directed weight editing without the same
gradient-based insertion procedure. Their targets include an MLP on
MNIST, ResNet-18 on CIFAR-10, and Qwen2.5-0.5B-Instruct on AG News.
These classification settings are not our structured-output agent
pipeline. The comparison concerns how edits are selected and which tasks
are measured, not a common benchmark establishing one method's
superiority.

\textbf{Weight-space detection and benchmarks.} Spectral-statistic
detection for LoRA adapters {[}40{]} and BackdoorLLM {[}41{]} supply
relevant defensive and benchmarking context. Their artifact types,
training/control data, and endpoints differ from this development
cohort. Reference-based weight differences, reference-free learned
detectors, and contextual forensic investigation answer related but
nonidentical questions.

\textbf{Our contribution in this context.} The value lies in the
eighteen-checkpoint/fourteen-family experimental breadth, the observed
sparse Phi-2 result with bounded reference metrics, the conditional
three-class taxonomy, exploratory PPO comparison, and recovery of
interpretable forensic evidence with explicit benign-control
limitations. We do not claim priority for weight editing, a proven
minimum sparsity floor, unique access to payload evidence, or
universally superior detection.

\subsection{Adversarial Machine
Learning}\label{adversarial-machine-learning}

Adversarial examples {[}10,11,12{]}, universal triggers {[}27{]},
jailbreak suffixes {[}13{]}, and prompt-injection studies {[}14,15{]}
manipulate model inputs under their evaluated conditions. Tool-use
backdoor research {[}32{]} further motivates separating generated action
content from downstream effects. A weight edit changes the artifact used
for subsequent inference, whereas an input attack supplies particular
runtime content; neither distinction guarantees persistence through
every transformation or success in every context.

Stable-ID insertion (§4) avoids a specific retokenization ambiguity, not
contextual dependence. Likewise, low reference-prefix drift in our Phi-2
artifact does not imply invisibility to input defenses or
complete-answer preservation. Artifact inspection and deployment
authorization address different parts of the threat and can be evaluated
together without claiming that either universally subsumes the other.

\subsection{Model Editing and Knowledge
Neurons}\label{model-editing-and-knowledge-neurons}

ROME {[}16{]}, MEMIT {[}17{]}, and knowledge-neuron analysis {[}18{]}
investigate editing and localization of model associations. Conditional
target insertion shares the idea of changing weights to influence
outputs, but differs in objective, constraints, and evaluation. Our
magnitude-dormancy statistic is not a test of functional inactivity or a
causal knowledge-neuron mechanism. Neither its counts nor the observed
response classes establish which circuit property causes a particular
insertion outcome.

\subsection{RL for Exploration and LLM
Testing}\label{rl-for-exploration-and-llm-testing}

Deep RL {[}19{]}, trust-region and proximal policy optimization
{[}20,21{]}, and novelty search {[}22{]} motivate evaluating learned
exploration under a specified reward. FuzzLLM {[}23{]} and PromptBench
{[}24{]} study input-space testing. Our PPO experiment instead selects
weight perturbations and measures finite-reference response per nominal
path cost. The four-model two-sided Welch comparison shows descriptive
search variation, not superiority over random exploration, equivalence,
or controllable failure across all inputs.

\subsection{Data-Poisoning Backdoors and Memory-Leak
Detection}\label{data-poisoning-backdoors-and-memory-leak-detection}

Sleeper Agents {[}8{]} and memory-leak detection {[}37{]} motivate
testing whether characteristic text can be recovered through inference.
Our local study (§7.11) reports one FFN fixture, one data-poison
fixture, and a benign adaptation with different structural and leak
observations. It does not establish statistical indistinguishability of
all weight distributions, universal memorization by data-poisoned
models, or mutually exclusive detector coverage by attack class. These
bounded cross-tests motivate examining both channels under matched
protocols and benign controls.

\subsection{Model Provenance and Lineage
Attribution}\label{model-provenance-and-lineage-attribution}

Model-provenance tools such as Cisco's Model Provenance Kit {[}38{]}
inspect metadata, tokenization, and weight-derived fingerprints to
support lineage attribution. Attribution is not a certificate of benign
behavior. Conversely, an anomaly or unusual output does not alone
identify an artifact's origin or malicious intent. We do not report a
matched evaluation showing that a particular provenance tool misses
every sparse edit. Signing, lineage analysis, weight comparison, and
contextual behavior provide different evidence that must be interpreted
within their measured scope.

\section{Ethical Consideration}\label{ethical-consideration}

\subsection{Dual-Use Nature}\label{dual-use-nature}

Our methods are dual-use: techniques for characterizing weight-space
vulnerabilities can also support exploitation. We address this tension
through responsible security research practices established in
adversarial ML {[}11,12{]} and vulnerability disclosure.

\textbf{Bounded reporting.} We separate short-marker effects, structured
output, parsed action content, and executed task impact. The scanner's
contextual probing does not dispatch the requests it parses. Historical
partial and unsuccessful records remain visible rather than being
presented as completed successes. The paper does not establish that
every architecture, trigger, or deployment is susceptible under one
procedure.

\textbf{Defensive purpose and deployment limits.} The intended value is
evidence for model investigation: identifying differences, candidate
inputs, and interpretable output content while documenting
benign-control failures. These observations motivate further defense
development, but do not validate immediate registry admission,
calibrated pruning thresholds, or a zero-cost universal mitigation.

\subsection{Broader Impact}\label{broader-impact}

The model supply chain will continue to grow as open-weight models
proliferate. We believe that characterizing the threat --- with
quantified attack difficulty, stealth metrics, and defense effectiveness
--- is strictly preferable to leaving the vulnerability undocumented.
The alternative (security through obscurity) has been consistently
rejected by the security community since Kerckhoffs' principle (1883).

\begin{center}\rule{0.5\linewidth}{0.5pt}\end{center}

\section{Limitations and Future Work}\label{limitations-and-future-work}

\subsection{Current Limitations}\label{current-limitations}

\textbf{Scope of models and attack.} The eighteen-checkpoint roster
spans dense transformers, MoE, state-space, and recurrent models, with
incomplete and reconstructed records explicitly identified. Eligible
edit matrices and trigger contexts vary. FFN-targeted non-transformer
runs do not test every state-update or gating intervention. The
three-class taxonomy is therefore conditional on the evaluated
procedures and must not be interpreted as architecture-wide immunity or
susceptibility.

\textbf{Compute, precision, and record completeness.} Local and cloud
campaigns use different hardware, execution paths, precision, and
budgets. Layerwise gradient computation and dequantization affect cost
and the object being evaluated. The gpt-oss bfloat16 campaigns are not a
security evaluation of its native MXFP4 deployment; quantized
representations do not make all gradient-based or surrogate optimization
impossible. Some cloud instability records are retained, but the six
primary paired drift/defense records do not cover the full roster.
Missing cells denote their stated evidence scope, not a universal
absence of results or an automatic failure.

\textbf{Statistical regime and triggers.} Insertion outcomes are
single-campaign observations; greedy action selection does not eliminate
sensitivity to numerical precision, tokenizer conventions, or
configuration. Four selected Gemma-2 triggers include three exact events
and one subunit result, not a population success-rate estimate. The
separate PPO-versus-random comparison has five seeds per arm and no
significant result under the two-sided Welch tests. Its nominal path
cost, precision limitations, clamped row indices, and differing action
distributions limit interpretation. These observations do not establish
equivalence or calibrated vulnerability ranking. Multi-token and broader
semantic-trigger coverage remain limited.

\textbf{Adaptive attempts: completed outcomes versus failures.} The
retained constructions include a z-clamp (\(z\leq2.99\)), dispersion
across top-ranked rows, a spectral-concentration penalty, and combined
constraints. Nine completed campaign summaries support the values below.
They are legacy scores, not calibrated success probabilities or proof of
robustness against an entire adaptive-attack class. Unconstrained values
come from separate selected baseline campaigns, not a controlled
effect-size experiment.

\begin{table*}[tb]\centering\caption{Nine completed adaptive campaign outcomes with separate
unconstrained baselines.}
\begin{adjustbox}{max width=\textwidth}\footnotesize\begin{tabular}[]{@{}
  >{\raggedright\arraybackslash}p{(\linewidth - 10\tabcolsep) * \real{0.1667}}
  >{\raggedright\arraybackslash}p{(\linewidth - 10\tabcolsep) * \real{0.1667}}
  >{\raggedright\arraybackslash}p{(\linewidth - 10\tabcolsep) * \real{0.1667}}
  >{\raggedright\arraybackslash}p{(\linewidth - 10\tabcolsep) * \real{0.1667}}
  >{\raggedright\arraybackslash}p{(\linewidth - 10\tabcolsep) * \real{0.1667}}
  >{\raggedright\arraybackslash}p{(\linewidth - 10\tabcolsep) * \real{0.1667}}@{}}

\toprule\noalign{}
\begin{minipage}[b]{\linewidth}\raggedright
Model
\end{minipage} & \begin{minipage}[b]{\linewidth}\raggedright
Unconstr.
\end{minipage} & \begin{minipage}[b]{\linewidth}\raggedright
z-clamp
\end{minipage} & \begin{minipage}[b]{\linewidth}\raggedright
Multilayer-K8
\end{minipage} & \begin{minipage}[b]{\linewidth}\raggedright
Spectral-λ0.1
\end{minipage} & \begin{minipage}[b]{\linewidth}\raggedright
Combined
\end{minipage} \\
\midrule\noalign{}

Phi-2 (2.78B) & 1.0 & --- & 0.000454748 & 0.000451211 & 0.000454748 \\
Llama-3.2-3B (3.21B) & 1.0 & --- & 0.000261190 & 0.000261190 &
0.000261190 \\
Qwen3-8B & 1.0 & 3.93e-7 & --- & --- & --- \\
gpt-oss-20b & 1.0 (α=0.10) & 0.001169510 & --- & --- & --- \\
Qwen2.5-7B & 0.002 & 5.72e-7 & --- & --- & --- \\
\bottomrule\noalign{}
\end{tabular}\end{adjustbox}\end{table*}

The three Phi-2 adaptive records contain 2,000 executed steps each; the
other six records contain 5,000 each. The retained Mistral adaptive log
ends with CUDA OOM on a cloud device and supplies no completed outcome
for this table. It is not a measured zero score, a successful defense
trial, or proof that the model cannot be attacked within a particular
memory capacity. Other blank cells are unevaluated here, not failures.

In particular, Qwen3-8B's \(3.93\times10^{-7}\) is the loss proxy
\(H(16.75)\), not token accuracy. None of the nine completed summaries
records an exact target event, but score ratios do not quantify
reductions in attack-success probability. These outcomes do not cover
every perturbation budget, target scope, quantization-aware
construction, multi-round attacker, or calibration-informed strategy.
Interrupted and missing evaluations remain outside the completed-result
denominator.

\textbf{Selected null-space and model-editing constructions.} A local
STEEREDIT-inspired construction {[}39{]} reports a small
clean-activation change, approximately \(8\times10^{-5}\) in its
normalized measure, alongside row z-score 31.9 and rank-1 ratio 1.000. A
local BadEdit/ROME-style GPT-2 construction {[}36{]} reports an exact
target event, z-score 16.95, and rank-1 ratio 1.000. These support the
observation that the tested low-rank edits can remain visible on a
reference-weight channel. They are not replications of the external
authors' full attack/defense benchmarks or proof that all
activation-stealthy or model-editing attacks are detected. The
model-editing fixture receives SUSPICIOUS for its non-action target,
illustrating the difference between a structural flag and the scanner's
action-based strongest verdict. Benign low-rank adaptation remains a
competing source of such signals.

\textbf{Blind scanner scope.} The 22-artifact panel is
development-exposed and includes related positive variants. It reports
four BACKDOORED and one SUSPICIOUS result among positives, seventeen
SUSPICIOUS clean results, and no CLEAN admissions. This is not a
validated no-false-admission guarantee or general clean/backdoored
classifier. The unresolved positive remains part of the denominator, and
runtime is not established as its sole causal explanation. Reference
access, context coverage, decoding, parser behavior, task authorization,
and benign adaptation all affect the operational question. Future
comparisons must match those conditions rather than transplanting
published detector percentages into this cohort.

\subsection{Future Work}\label{future-work}

\textbf{Expanded architecture evaluation.} Matched protocols, multiple
triggers, and architecture-appropriate target matrices would test how
stable the three observed regimes are. Completing partial
non-transformer campaigns and evaluating additional dense, sparse, and
hybrid models would extend coverage without presupposing a causal
architecture classification.

\textbf{Adaptive attack-defense evaluation.} Future work should separate
detector-development artifacts from evaluation artifacts, specify
adversary information and budgets prospectively, and retain failures and
unavailable outcomes. The nine completed adaptive records do not
establish an equilibrium defense strength or robustness to every
constrained optimization strategy.

\textbf{Registry deployment studies.} Forensic reports could support
human investigation of uploaded artifacts, but admission decisions
require measured benign-review burden, false alarms, missed positives,
runtime coverage, and handling of unresolved cases. The current
development cohort does not establish production readiness.

\textbf{Task-specific exploration.} PPO remains a possible experimental
approach for other well-defined objectives. A safety or capability
endpoint would need its own task, controls, reward validation, and
independent test. Maximizing the present finite-reference ratio does not
automatically produce a validated safety-critical exploit.

\textbf{Systematic scanner validation.} Priorities are independent
held-out lineages, legitimate adaptations, paired native-runtime tests,
complete action traces, and task-conditioned authorization checks.
Head-to-head comparisons should match artifact type, reference access,
query budget, and endpoint; reference-based, learned reference-free, and
contextual detectors should not be compared through incompatible
headline percentages. Larger and more diverse cohorts can test
generalization, but operational reliability remains a hypothesis to
evaluate.

\begin{center}\rule{0.5\linewidth}{0.5pt}\end{center}

\section{Conclusion}\label{conclusion}

We study post-training weight-space security through gradient-guided
insertion, exploratory PPO search, and model forensics. The evaluation
spans \textbf{eighteen checkpoints and fourteen architecture families},
including state-space and recurrent models, with completed, archival,
partial, and reconstructed records distinguished by status.

\textbf{Insertion and conditional taxonomy.} The Phi-2 artifact combines
an exact short-marker event, a small footprint, and low finite-reference
drift. Separate campaigns demonstrate structured-output feasibility.
\textbf{Gradient-Permeable, Scale-Threshold, and Gradient-Fragmented}
organize the configuration-dependent responses, including Gemma's
trigger sensitivity and qualified non-transformer records.

\textbf{Exploratory PPO evidence.} Recorded instability scores span
approximately 112,000-fold. PPO's higher Phi-2 mean is not significant
under the two-sided Welch test (\(p=0.082870\)), and none of the four
comparisons establishes a significant advantage.

\textbf{Forensic contribution.} Reference comparisons expose
modifications while contextual probes recover candidate and payload
evidence. The 22-artifact cohort contains four BACKDOORED positives, one
SUSPICIOUS positive, and seventeen SUSPICIOUS clean artifacts. Benign
adaptations and cross-defense cases identify where structural and
behavioral channels provide different evidence.

\textbf{Scope.} These configuration-specific results support further
artifact-level and deployment-aware auditing, with the finite-reference,
record-completeness, and development-cohort boundaries stated in the
body. General clean admission, task-level compromise, causal
architectural explanations, and universal mitigation remain questions
for independent evaluation.

\begin{center}\rule{0.5\linewidth}{0.5pt}\end{center}

\appendix

\subsection*{Master Record Index}\label{master-record-index}
\addcontentsline{toc}{subsection}{Master Record Index}

Table XXI defines the C/H/T/M campaign records and D measurement pairs.
Table links indicate where their results are used, excluding roster-only
appearances. The five-seed PPO study and adaptive comparisons are
separate evaluations.

\begin{table*}[tb]\centering\caption{Master index of campaign records and evaluated artifact
pairs.}
\begin{adjustbox}{max width=\textwidth}\footnotesize\begin{tabular}[]{@{}
  >{\raggedright\arraybackslash}p{(\linewidth - 8\tabcolsep) * \real{0.0833}}
  >{\raggedright\arraybackslash}p{(\linewidth - 8\tabcolsep) * \real{0.2583}}
  >{\raggedright\arraybackslash}p{(\linewidth - 8\tabcolsep) * \real{0.2417}}
  >{\raggedright\arraybackslash}p{(\linewidth - 8\tabcolsep) * \real{0.2417}}
  >{\raggedright\arraybackslash}p{(\linewidth - 8\tabcolsep) * \real{0.1750}}@{}}

\toprule\noalign{}
\begin{minipage}[b]{\linewidth}\raggedright
ID
\end{minipage} & \begin{minipage}[b]{\linewidth}\raggedright
Model
\end{minipage} & \begin{minipage}[b]{\linewidth}\raggedright
Campaign / measurement
\end{minipage} & \begin{minipage}[b]{\linewidth}\raggedright
Record status
\end{minipage} & \begin{minipage}[b]{\linewidth}\raggedright
Used in tables
\end{minipage} \\
\midrule\noalign{}

C01 & Phi-2 & Short marker & Completed & VI, VII, IX, XV \\
C02 & Llama-3.2-3B & Short marker & Completed & VII, IX, XV \\
C03 & Gemma-2-2B & Short marker & Completed & VII, VIII, IX, XV \\
C04 & Phi-4-mini & Short marker & Completed & VI, VII, XV \\
C05 & DeepSeek-R1-1.5B & Short marker & Completed & VII, VIII, XV \\
C06 & Qwen2.5-3B & Short marker & Completed & VII, VIII, XV \\
C07 & StableLM-1.6B & Short marker & Completed & VII, VIII, IX, XV \\
C08 & OLMo-2-7B & Short marker & Completed & VII, VIII, XV \\
C09 & Mistral-7B & Short marker & Completed & VII, IX, XV \\
C10 & DeepSeek-R1-Llama-8B & Short marker & Completed & VII, IX, XV \\
C11 & Qwen3-8B & Short marker & Completed & VII, IX, XV \\
C12 & gpt-oss-20b & Short marker & Completed & VII, VIII, IX, XV \\
C13 & Llama-3.1-8B & Short marker & Completed & VII, XV \\
C14 & Qwen2.5-7B & Short marker & Completed; archived JSON & VII, IX,
XV \\
C15 & Gemma-4-31B & Short-marker summary & Reconstructed & VII, XV \\
C16 & RWKV-7-2.9B & Structured probe & Completed & VII, VIII, XV \\
H01 & GPT-2 & Short marker & Archival & VI, VII, IX, XV \\
H02 & GPT-2 & Structured request & Archival & IX, XV \\
T01 & Phi-2 & Structured request & Completed & IX, XV \\
T02 & Llama-3.2-3B & Structured request & Completed & IX, XV \\
T03 & Gemma-2-2B & Structured request & Completed & IX, XV \\
T04 & StableLM-1.6B & Structured request & Completed & IX, XV \\
T05 & Mistral-7B & Structured request & Completed & IX, XV \\
T06 & DeepSeek-R1-Llama-8B & Structured request & Completed & IX, XV \\
T07 & Qwen3-8B & Structured request & Completed & IX, XV \\
T08 & Qwen2.5-7B & Structured request & Completed; archived JSON & IX,
XV \\
T09 & gpt-oss-20b & Structured request & Completed & IX, XV \\
M01 & Mamba-2-2.7B & Short-target log & Partial & VII, XV \\
M02 & Mamba-2-2.7B & Structured-target log & Partial & XV \\
D01 & Phi-2 & Drift / rounding / defense & Evaluated; exact baseline &
III, X, XV \\
D02 & Llama-3.2-3B & Drift / rounding / defense & Evaluated; exact
baseline & X, XV \\
D03 & Gemma-2-2B & Drift / rounding / defense & Evaluated; unmatched
baseline & X, XV \\
D04 & Phi-4-mini & Drift / rounding / defense & Evaluated; exact
baseline & X, XV \\
D05 & DeepSeek-R1-1.5B & Drift / rounding / defense & Evaluated;
unmatched baseline & X, XV \\
D06 & Qwen2.5-3B & Drift / rounding / defense & Evaluated; unmatched
baseline & X, XV \\
\bottomrule\noalign{}
\end{tabular}\end{adjustbox}\end{table*}

\textbf{Status and selection.} Completed means a normal reported stop,
including budget exhaustion; archival means a historical report with
limited primary evidence; partial means a log without campaign
completion; reconstructed means a recovered summary. Selection follows
named local baselines, the cloud-v2 collection, and archived Qwen2.5-7B
records rather than the best of alternative runs. H01/H02 model
attribution comes from contemporaneous notes; C15 was recovered after
disk-full and its timing is unused.

\textbf{Pairing and missing fields.} D records identify saved artifact
pairs rather than necessarily the states at a campaign maximum.
Llama-3.1's 2,253-step demonstration is separate from C13. C01 records
the 300-step, 332.078-second Phi-2 event; D01 supplies its
6,092/2,779,683,840 changed entries, L2 distance 0.3462730724, and
maximum absolute difference 0.0936737061. Original trigger IDs, target
arrays, templates, and runtime fields are not uniformly retained; later
reconstructions do not retroactively authenticate them.

\subsection*{Reference-Optional Scanner
Cohort}\label{reference-optional-scanner-cohort}
\addcontentsline{toc}{subsection}{Reference-Optional Scanner Cohort}

The 22-artifact development cohort contains five positive fixtures and
seventeen clean artifacts. The positives include Phi-4-mini,
Llama-3.1-8B, and related Llama-3B embedding and packaging variants.
Four positives receive BACKDOORED; the Ollama-related variant and all
seventeen clean artifacts receive SUSPICIOUS. None receives CLEAN.
Variant names do not by themselves certify stored tensor formats.

These reports informed development, and the related Llama packaging
variants are not independent attack lineages. Recorded times span
6.6-145.6 seconds. Duplicate rechecks and an attempt without a completed
verdict are not counted as independent outcomes. Most summaries contain
lab-v2 results with the earlier branch skipped. A separate decision
replay matched six available full reports; the canonical summaries lack
the complete phase fields required for that strict replay. These
record-level checks are not independent detector validation.

\subsection*{Additional Measurement
Boundaries}\label{additional-measurement-boundaries}
\addcontentsline{toc}{subsection}{Additional Measurement Boundaries}

Table VIII's six transformer scale series compare the listed default,
0.10, and 0.15 configurations. RWKV uses different settings: 5,000 steps
at 0.15 and a completed 10,000-step episode followed by an incomplete
episode at 0.20 under a later float16/five-token configuration. C16 is a
separate probe. M01/M02 contain 108/774 partial steps with maximum
displayed scores 0.0040/0.2733.

The Gemma trigger study uses three multi-tab anchors and one
non-whitespace token under the declared common scale and budget. Its
selected scores are approximately 1.000, 0.624, 1.000, and 1.000. The
separate joint-cap comparison changes both attention-logit and
final-logit settings, records 3,000 ON steps and 2,654 OFF steps, and
does not retain modified checkpoints for independent endpoint replay. It
does not isolate the individual effects of the two caps.

The PPO comparison contains forty observations: four checkpoints, two
search methods, and five seeds per arm. The single-campaign instability
summaries plotted in Fig. 5 are separate; their best score and best
maximum KL need not occur at the same state. The nine completed adaptive
summaries in §11.1 are also separate from the Mistral OOM attempt. None
of those incomplete or unavailable results is assigned a successful
defense outcome.

The local cross-defense study in §7.11 uses one FFN-edit fixture, one
data-poison sleeper, and one benign adaptation. Training metadata
specifies LoRA r16 merged models, 1,299 benign examples versus 2,599
sleeper examples, and three epochs each. The 31/39 target-string count,
0/39 trigger count, and 0.8205 motif score measure different endpoints.
A modified-tensor parameter fraction is not an exact changed-scalar
count, and one benign counterexample does not establish population-level
performance or statistical equivalence.

\begin{center}\rule{0.5\linewidth}{0.5pt}\end{center}

\section*{References}\label{references}
\addcontentsline{toc}{section}{References}

{[}1{]} M. Zalewski, ``American Fuzzy Lop (AFL),'' 2014. {[}Online{]}.
Available: https://lcamtuf.coredump.cx/afl/

{[}2{]} K. Serebryany, ``LibFuzzer: A library for coverage-guided fuzz
testing,'' LLVM Project, 2015.

{[}3{]} M. Böhme, V.-T. Pham, and A. Roychoudhury, ``Coverage-based
greybox fuzzing as Markov chain,'' in \emph{Proc. ACM CCS}, 2016,
pp.~1032-1043.

{[}4{]} K. Pei, Y. Cao, J. Yang, and S. Jana, ``DeepXplore: Automated
whitebox testing of deep learning systems,'' in \emph{Proc. ACM SOSP},
2017, pp.~1-18.

{[}5{]} T. Gu, B. Dolan-Gavitt, and S. Garg, ``BadNets: Identifying
vulnerabilities in the machine learning model supply chain,''
arXiv:1708.06733, 2017.

{[}6{]} Y. Liu, S. Ma, Y. Aafer, W.-C. Lee, J. Zhai, W. Wang, and X.
Zhang, ``Trojaning attack on neural networks,'' in \emph{Proc. NDSS},
2018.

{[}7{]} B. Wang, Y. Yao, S. Shan, H. Li, B. Viswanath, H. Zheng, and B.
Y. Zhao, ``Neural Cleanse: Identifying and mitigating backdoor attacks
in neural networks,'' in \emph{Proc. IEEE S\&P}, 2019, pp.~707-723.

{[}8{]} E. Hubinger et al., ``Sleeper Agents: Training deceptive LLMs
that persist through safety training,'' arXiv:2401.05566, 2024.

{[}9{]} A. Rakin, Z. He, and D. Fan, ``Bit-Flip Attack: Crushing neural
network with progressive bit search,'' in \emph{Proc. ICCV}, 2019.

{[}10{]} C. Szegedy, W. Zaremba, I. Sutskever, J. Bruna, D. Erhan, I.
Goodfellow, and R. Fergus, ``Intriguing properties of neural networks,''
in \emph{Proc. ICLR}, 2014.

{[}11{]} I. Goodfellow, J. Shlens, and C. Szegedy, ``Explaining and
Harnessing Adversarial Examples,'' in \emph{Proc. ICLR}, 2015.

{[}12{]} A. Madry, A. Makelov, L. Schmidt, D. Tsipras, and A. Vladu,
``Towards deep learning models resistant to adversarial attacks,'' in
\emph{Proc. ICLR}, 2018.

{[}13{]} A. Zou, Z. Wang, J. Z. Kolter, and M. Fredrikson, ``Universal
and transferable adversarial attacks on aligned language models,''
arXiv:2307.15043, 2023.

{[}14{]} K. Greshake, S. Abdelnabi, S. Mishra, C. Endres, T. Holz, and
M. Fritz, ``Not what you've signed up for: Compromising real-world
LLM-integrated applications with indirect prompt injection,'' in
\emph{Proc. AISec@CCS}, 2023.

{[}15{]} F. Perez and I. Ribeiro, ``Ignore Previous Prompt: Attack
Techniques For Language Models,'' arXiv:2211.09527, 2022.

{[}16{]} K. Meng, D. Bau, A. Andonian, and Y. Belinkov, ``Locating and
editing factual associations in GPT,'' in \emph{Proc. NeurIPS}, 2022.

{[}17{]} K. Meng, A. S. Sharma, A. Andonian, Y. Belinkov, and D. Bau,
``Mass-Editing Memory in a Transformer,'' in \emph{Proc. ICLR}, 2023.

{[}18{]} D. Dai, L. Dong, Y. Hao, Z. Sui, B. Chang, and F. Wei,
``Knowledge neurons in pretrained transformers,'' in \emph{Proc. ACL},
2022.

{[}19{]} V. Mnih et al., ``Playing Atari with Deep Reinforcement
Learning,'' arXiv:1312.5602, 2013.

{[}20{]} J. Schulman, S. Levine, P. Abbeel, M. Jordan, and P. Moritz,
``Trust region policy optimization,'' in \emph{Proc. ICML}, 2015.

{[}21{]} J. Schulman, F. Wolski, P. Dhariwal, A. Radford, and O. Klimov,
``Proximal Policy Optimization Algorithms,'' arXiv:1707.06347, 2017.

{[}22{]} J. Lehman and K. O. Stanley, ``Abandoning objectives: Evolution
through the search for novelty alone,'' \emph{Evolutionary Computation},
vol.~19, no. 2, pp.~189-223, 2011.

{[}23{]} D. Yao, J. Zhang, I. G. Harris, and M. Carlsson, ``FuzzLLM: A
Novel and Universal Fuzzing Framework for Proactively Discovering
Jailbreak Vulnerabilities in Large Language Models,'' in \emph{Proc.
IEEE ICASSP}, 2024. arXiv:2309.05274.

{[}24{]} K. Zhu et al., ``PromptBench: Towards evaluating the robustness
of large language models on adversarial prompts,'' arXiv:2306.04528,
2023.

{[}25{]} N. Elhage et al., ``A Mathematical Framework for Transformer
Circuits,'' Transformer Circuits, 2021.

{[}26{]} N. Elhage et al., ``Toy Models of Superposition,'' Anthropic,
2022.

{[}27{]} E. Wallace, S. Feng, N. Kandpal, M. Gardner, and S. Singh,
``Universal Adversarial Triggers for Attacking and Analyzing NLP,'' in
\emph{Proc. EMNLP}, 2019.

{[}28{]} K. Kurita, P. Michel, and G. Neubig, ``Weight Poisoning Attacks
on Pre-trained Models,'' in \emph{Proc. ACL}, 2020, pp.~2793-2806.

{[}29{]} S. Marks and M. Tegmark, ``The Geometry of Truth: Emergent
Linear Structure in Large Language Model Representations of True/False
Datasets,'' in \emph{Proc. Conference on Language Modeling (COLM)},
2024. arXiv:2310.06824.

{[}30{]} A. Zou, L. Phan, S. Chen, J. Campbell, et al., ``Representation
Engineering: A Top-Down Approach to AI Transparency,'' arXiv:2310.01405,
2023.

{[}31{]} M. Abu Baker and L. Babu-Saheer, ``Mechanistic Exploration of
Backdoored Large Language Model Attention Patterns,'' arXiv:2508.15847,
2025.

{[}32{]} W. Zhang and S. Pei, ``Your LLM Agent Can Leak Your Data: Data
Exfiltration via Backdoored Tool Use,'' arXiv:2604.05432; in \emph{Proc.
ACL}, 2026.

{[}33{]} Sigstore, ``Sigstore Documentation.'' {[}Online{]}. Available:
\url{https://docs.sigstore.dev/}

{[}34{]} Hugging Face, ``Signing commits with GPG.'' {[}Online{]}.
Available: \url{https://huggingface.co/docs/hub/security-gpg}

{[}35{]} S. Hong, N. Carlini, and A. Kurakin, ``Handcrafted Backdoors in
Deep Neural Networks,'' in \emph{Proc. NeurIPS} (Oral), 2022.
arXiv:2106.04690.

{[}36{]} Y. Li, T. Li, K. Chen, J. Zhang, S. Liu, W. Wang, T. Zhang, and
Y. Liu, ``BadEdit: Backdooring large language models by model editing,''
in \emph{Proc. ICLR}, 2024. arXiv:2403.13355.

{[}37{]} B. Bullwinkel, G. Severi, K. Hines, A. Minnich, R. S. S. Kumar,
and Y. Zunger, ``The Trigger in the Haystack: Extracting and
Reconstructing LLM Backdoor Triggers,'' arXiv:2602.03085, 2026.

{[}38{]} E. Aghaei, A. Chang, A. Garg, and S. Mendapara, ``Introducing
Model Provenance Kit: Know Where Your AI Models Come From,'' Cisco AI
Defense, Apr.~2026. {[}Online{]}. Available:
https://github.com/cisco-ai-defense/model-provenance-kit

{[}39{]} R. Yin, T. Han, N. Xu, C. Li, P. He, C. Zhou, J. Wang, Z. Fu,
T. Du, J. Li, and S. Ji, ``Compiling Activation Steering into Weights
via Null-Space Constraints for Stealthy Backdoors,'' in \emph{Proc. ACL
(Long Papers)}, 2026. arXiv:2604.12359.

{[}40{]} D. Puertolas Merenciano, E. Vasyagina, K. Zhu, J. Ferrando, and
M. Chaudhary, ``Weight-Space Detection of Backdoors in LoRA Adapters,''
in \emph{Proc. ICLR}, 2026. arXiv:2602.15195.

{[}41{]} Y. Li, H. Huang, Y. Zhao, X. Ma, and J. Sun, ``BackdoorLLM: A
Comprehensive Benchmark for Backdoor Attacks and Defenses on Large
Language Models,'' in \emph{Proc. NeurIPS, Datasets and Benchmarks
Track}, 2025. arXiv:2408.12798.

{[}42{]} Y. Zeng, G. Severi, B. Bullwinkel, E. Horvitz, and D.
Papailiopoulos, ``Lazare, veni foras: Reprogramming Dormant Neurons to
Backdoor AI Agents,'' arXiv:2611.15118v1, 11 November 2026.

{[}43{]} Z. Wang, J. Lu, and Z. Li, ``Can LLMs Write Backdoors? An
Empirical Security Study of Weight-Space Programming,''
arXiv:2611.17395v1, 12 November 2026.

{[}44{]} B. Tran, J. Li, and A. Madry, ``Spectral Signatures in Backdoor
Attacks,'' in \emph{Proc. NeurIPS}, 2018.

{[}45{]} K. Liu, B. Dolan-Gavitt, and S. Garg, ``Fine-Pruning: Defending
Against Backdooring Attacks on Deep Neural Networks,'' in \emph{Proc.
RAID}, 2018. arXiv:1805.12185.

\end{document}